\documentclass[a4paper,12pt]{article}

\usepackage{amsmath,amsfonts,amssymb,epsfig}

\numberwithin{equation}{section}

\usepackage{cancel,comment}
\usepackage{cite}

\def\A{\mathcal{A}}

\def\2b2[#1,#2][#3,#4]{\left( \begin{array}{cc} #1 & #2 \\ #3 & #4 \end{array} \right)}
\def\3b3[#1,#2,#3][#4,#5,#6][#7,#8,#9]{\left( \begin{array}{ccc} #1 & #2 #3 \\ #4 & #5 & #6\\#7&#8&#9\end{array} \right)}
\def\6diag[#1,#2,#3]{\left( \begin{array}{cccccc} #1 & 0 & 0 & 0 & 0 & 0 \\ 0 & #1 & 0 & 0 & 0 & 0 \\ 0 & 0& #2 & 0 & 0 &0 \\ 0 & 0 & 0 & #2 & 0 & 0 \\ 0 & 0 & 0 & 0 & #3 & 0 \\ 0 & 0 & 0 & 0 & 0 & #3  \end{array} \right)}
\def\3diag[#1,#2,#3]{\left(\begin{array}{ccc} #1 & 0 & 0 \\ 0 & #2 &0\\ 0&0&#3 \end{array}\right)}

\def\V{\mathcal{V}}

\newcommand{\re}{\mathrm{Re\;}}
\newcommand{\half}{\frac{1}{2}}
\def\vt{\vartheta_1}
\def\bra{\langle}
\def\ket{\rangle}

\def\tr{\mathrm{tr}}
\def\ap{\alpha^{\prime}}

\def\beq{\begin{equation}}
\def\eeq{\end{equation}}
\def\bal{\begin{align}}
\def\eal{\end{align}}
\def\nn{\nonumber}
\def\mr{\mathrm}
\newcommand{\bea}{\begin{eqnarray}}
\newcommand{\eea}{\end{eqnarray}}

\newcommand{\C}[1]{\mathcal{#1}}

\def\ov{\overline}

\author{
P. Anastasopoulos$^{1}$\footnote{pascal@hep.itp.tuwien.ac.at},~
I. Antoniadis$^{2}$\footnote{ignatios.antoniadis@cern.ch}~\footnote{On leave from CPHT (UMR CNRS 7644) Ecole Polytechnique, F-91128 Palaiseau},~
\\ K.~Benakli$^3$\footnote{kbenakli@lpthe.jussieu.fr}~,
M.~D.~Goodsell$^4$\footnote{mark.goodsell@desy.de},~
A.~Vichi$^5$\footnote{alessandro.vichi@epfl.ch}}
\date{}
\title{\vspace{-3cm}
\hfill{\small{CERN-PH-TH/2011-095}}\\\vspace{-0.5cm}
\hfill{\small{DESY 11-069}}\\\vspace{-0.5cm}
\hfill{\small{TUW-11-09}}\\[2cm]
{One-loop adjoint masses for non-supersymmetric intersecting branes}}
\begin{document}
\maketitle
\vspace{-1cm}
\begin{center}
\emph{$^1$ Technische Univ. Wien Inst. fur Theoretische Physik, A-1040 Vienna, Austria\\
$^2$ Department of Physics, CERN Theory Division, CH-1211, Geneva 23, Switzerland\\
$^3$ Laboratoire de Physique Th\'eorique et Hautes Energies, CNRS, UPMC Univ Paris 06
Boite 126, 4 Place Jussieu, 75252 Paris cedex 05, France \\
$^4$ Deutsches  Elektronen-Synchrotron, DESY, Notkestra\ss e 85, 22607  Hamburg, Germany\\
$^5$ Institut de Th\'eorie des Ph\'enom\'enes Physiques, EPFL, CH-1015 Lausanne, Switzerland}
\end{center}
\abstract{We consider breaking of supersymmetry in intersecting D-brane configurations by slight deviation of the angles from their supersymmetric values. We compute the masses generated by radiative corrections for the adjoint scalars on the brane world-volumes. In the open string channel, the string two-point function receives contributions only from the infrared and the ultraviolet limits. The latter is due to tree-level closed string uncanceled NS-NS tadpoles, which we explicitly reproduce from the effective Born-Infeld action. On the other hand, the infrared region reproduces the one-loop mediation of supersymmetry breaking in the effective gauge theory, via messengers and their Kaluza-Klein excitations. In the toroidal set-up considered here, it receives contributions only from  $N\approx 4$ and $N\approx 2$ supersymmetric configurations, and thus always leads at leading order to a tachyonic direction, in agreement with effective field theory expectations.}

\newpage

\section{Introduction}
\label{SEC:INTRODUCTION}

D-branes appear to be a powerful tool for engineering gauge theories upon their embedding in higher dimensional spaces. Of greatest importance for relating to the real world are  configurations with softly broken supersymmetric low energy effective field theories. A simple way to achieve such a breaking is to introduce a magnetic field  which, due to the different  couplings with the spins,  induces a mass splitting between fermions with different chiralities and with bosons \cite{Bachas:1995ik, Angelantonj:2000hi}. The same splitting can be mapped upon T-duality into  branes intersecting at angles \cite{Berkooz:1996km, Blumenhagen:2000wh}, providing a simple geometrical description. 

A supersymmetric vacuum can be obtained through a specific choice of intersection angles between D-branes.  Then, a breaking of supersymmetry with a size parametrically smaller than the string scale can  be obtained by choosing the angles (or the magnetic fluxes) slightly away from their supersymmetric values \cite{Antoniadis:2006eb, Antoniadis:2005em, Antoniadis:1997mm}. At tree-level, this  breaking appears as mass shifts in the spectrum of open strings localised at the brane intersections.  Through radiative corrections, the breaking is  communicated  to the other states living on the brane world-volume. We will carry out here an explicit computation of such effects. We will be particularly interested in the induced masses for  the adjoint representations of the gauge group. Indeed, it is known that this mechanism generates for instance one-loop Dirac gaugino masses, but some adjoint scalars tend to become tachyonic in the effective field theory, which is the main obstruction to building an interesting viable model of supersymmetry breaking.

We will perform the string computation in the case of toroidal compactifications (with or without orientifold and orbifold projections) as the world-sheet description by free fields  allows the straightforward use of conformal field theory techniques. Considering that the breaking through a magnetic field can be described  as the appearance of a non-vanishing $D$-term, we can then compute in the effective field theory the radiative masses generated on the world-volume. The  results depend on the number of supersymmetries that are originally preserved by the brane intersections before having the small shift in angles that induces supersymmetry breaking. The mass corrections vanish for an originally $N=1$ (written as $N \approx 1$ ) sector with non-vanishing intersection angles in the three tori. This is due to the absence of couplings between the messengers and  scalars in adjoint representations at the one-loop level. The  $N\approx2$ and $N\approx4$ cases  correspond to rotating by a small angle branes that are otherwise parallel in one and three  tori, respectively. In these simple cases, one can derive the one-loop effective potential and read from there the masses of the adjoint representations. These results will be reproduced explicitly through a string one-loop vacuum amplitude, and appear accompanied with similar (sub-leading) contributions from the Kaluza-Klein excitations. At leading order, the obtained mass matrix is traceless, and signals the presence of a tachyonic direction.

The string computation gives in addition a tree-level closed string divergence in the ultraviolet limit of the open string channel. 
We shall show how this is actually a reducible contribution, matching the expectations from supergravity in the presence of NS-NS tadpoles through the emission of a  massless  dilaton and internal metric moduli. These results are expected to be drastically modified when taking moduli stabilization into account, causing a shift in the vacuum of the theory and cancellation of the tadpoles.

Beyond expected field theory contributions, it is interesting to find that there is no extra contribution (at leading order in the supersymmetry breaking parameter expansion) from the massive string states due to the form of the correlation functions and the boundary conditions involved in the computation of the amplitude, a feature that needed an explicit check by writing down the two-point correlation functions.

%%%% SECTIONS 
% 1 \label{SEC:INTRODUCTION}
% 2 \label{SEC:SETUP} 
% 3 \label{SEC:CW} 
% 4 \label{SEC:CFT} 
% 5 \label{SEC:POTENTIAL} 
% 6 \label{SEC:TWOPOINT}
% 7 \label{SEC:STRINGY}
% 8 \label{SEC:NONSTRINGY}
% 9 \label{SEC:MASSESBEYONDLEADING}
% 10 \label{SEC:SUGRA}
% 11 \label{SEC:CONCLUSIONS}

The paper is organised as follows. In section \ref{SEC:SETUP} we describe a simple string set-up that allows to perform simultaneously both explicit string and effective field theory computations. In section \ref{SEC:CW} we explicitly compute the result from the one-loop effective potential within the low energy effective gauge theory. Section \ref{SEC:CFT} introduces some basic vertex operators, correlation and partition functions used later on. Section \ref{SEC:POTENTIAL}  shows the string derivation of the results of section \ref{SEC:CW}. Section \ref{SEC:TWOPOINT} discusses how masses can arise from string two-point functions. The contributions to the amplitude that cannot be determined from the effective potential are then explicitly derived in section \ref{SEC:STRINGY}, showing that they come purely from light (massless) closed string states. The effective potential results (arising from the open string channel) are reproduced via a string two-point function calculation in section \ref{SEC:NONSTRINGY}. Section \ref{SEC:MASSESBEYONDLEADING} investigates the ubiquitous presence of a tachyonic direction. The closed string contribution is exactly matched with the effective supergravity expectation in section \ref{SEC:SUGRA}. An appendix \ref{APPENDIX:FT} provides a detailed calculation of the field theory limit of the $N\approx 2$ case, showing how the different feynman diagram contributions arise from the string amplitude.

\section{The string set-up }
\label{SEC:SETUP}

In intersecting brane models on tori, adjoint fields arise as position and wilson line moduli of branes. We can determine their kinetic terms and couplings to closed string fields by examination of the Dirac-Born-Infeld action; indeed, in the case that the intersection of two branes preserves two or four supersymmetries, the latter determine the adjoint couplings to the  non-chiral states stretched between the branes. If we deform the intersection angles by a small amount then a mass is generated for some of these adjoints, which can be calculated in the low energy field theory. This can be done either by computing diagrams or by an effective potential calculation; we shall choose the latter, since the computation can be done purely from the spectrum. 

The background will be $\mathbb{T}^2_1 \times \mathbb{T}^2_2 \times \mathbb{T}^2_3$ with radii $R_{1}^i, R_{2}^i, i=1,2,3$. We then define the K\"ahler modulus of the torus to be $T^j = T_1^j + i T_2^j = i  R_1^j R_2^j \sin \alpha^j$, where $\alpha $ is the angle between the axes (we shall generally take it to be $\pi/2$ for simplicity). The complex structure is given by $\frac{R_2^j}{R_1^j} e^{i\alpha^j}$. $L_j$ is the length of the open strings, given in terms of the wrapping numbers $n_j, m_j$ by
\begin{align}
L_j ~=~& 2\pi \sqrt{\frac{T_2^j}{U_2^j}| n^j + m^j U^j|^2} ~=~ 2\pi \sqrt{(n^j)^2 (R_1^j)^2 + (m^j)^2 (R_2^j)^2 + 2 n^j m^j R_a^j R_2^j \cos \alpha^j}.
\end{align} 
We shall also need the quantity 
\beq
\V^j ~\equiv~ L^j/2\pi,
\eeq
which is an effective radial parameter; if the brane is aligned along one of the axes of a torus, then this is just the radius of the corresponding torus. The reason for the definition is that this is the quantity that appears in Kaluza-Klein momenta. 

Consider two branes $a$ and $b$ intersecting at angle $\pi \theta_{ab}^i$ in the $i^{th}$ torus, breaking supersymmetry by a small amount such that the angles obey
\beq
\sum_{i=1}^3 \theta_{ab}^i ~=~ 2\epsilon.
\eeq
If the angles are such that when $\epsilon=0$ they obey $\theta_{ab}^i \notin \{0,1\} \,\forall\, i$, this is an \emph{almost $N=1$ sector}, or $N \approx 1$ for short; in this case the adjoints do not have renormalizable couplings to matter fields and the field theory effective potential generates no mass for them. If $\theta_{ab}^i \in \{0,1\} \,\forall\, i$ this is an \emph{$N\approx 4$ sector}. If $\theta_{ab}^i \in \{0,1\} , \theta_{ab}^{j \ne i},\theta_{ab}^{k \ne j \ne i} \notin \{0,1\}$ we call this an \emph{$N\approx 2$ sector}. Here there are two cases: either $\theta_{ab}^i = 2\epsilon$ for some $i$, which must be treated similarly to an $N=1$ sector; or $\theta_{ab}^i \in \{0,1\}$. We shall define the intersection number 
\begin{align}
I_{ab} ~\equiv~& \prod_{\kappa | I^{\kappa}_{ab} \ne 0} I_{ab}^\kappa \nn\\
I_{ab}^\kappa ~\equiv~& (n_a^\kappa m_b^\kappa - n_b^\kappa m_a^\kappa).
\end{align}
where $n_a^\kappa, m_a^\kappa$ are the number of times that the $R_1^\kappa, R_2^\kappa$ cycles are wrapped in the $\kappa$ torus respectively. These are naturally related to the angles, for example via the identity that for $I_{ab}^\kappa \ne 0$
\begin{align}
\frac{\V_a^\kappa \V_b^\kappa}{ T_2^\kappa} ~=~ \frac{I_{ab}^\kappa}{\sin \pi \theta_{ab}^\kappa} ~=~\frac{|I_{ab}^\kappa|}{|\sin \pi \theta_{ab}^\kappa|} . 
\end{align}

\section{Adjoint scalar masses from the low energy theory}
\label{SEC:CW}

Here we present what can be calculated from a field theory point of view using the spectrum and the Coleman-Weinberg potential, before reproducing these results via string theory and then calculating the masses from ultra-violet effects. 

\subsection{$N\approx 4$}

This case is essentially $T$-dual to the model considered by \cite{Bachas:1995ik}. It has subsequently been used to study inflation in, for example, \cite{GomezReino:2002fs,Hebecker:2011hk}. We determine the effective potential by analyzing the spectrum of states stretched between the two branes. They fall into bifundamental representations $(\mathbf{1}_a,-\mathbf{1}_b)+(-\mathbf{1}_a,\mathbf{1}_b)$ of the $U(1)$s on each brane. The spectrum is given by three factors: 
\beq
M^2_n ~=~  M_0^2 + 2n|\epsilon|/\ap + \C{M}^2
\eeq
where $n=0, 1, 2,....$ denotes the number of pseudo-zero-mode operator insertions corresponding to the multiplicity of Landau levels (i.e. the bosonic operators $\alpha_0, \alpha_0^\dagger$ - in the limit $\epsilon=0$ the torus decompactifies and these become momentum modes); $\C{M}^2$ depends upon the Lorentz representation as given in table \ref{Napprox4spectrum}. Without loss of generality we take $\theta_{ab}^3 = 2\epsilon, \theta_{ab}^1 = \theta_{ab}^2 = 0$ and so there are three complex scalars $\Phi_i$, their would-be fermionic superpartners $\Psi_i$, a vector and gaugino. Then   $M_0^2 = \frac{y^2}{4\pi^2 (\ap)^2} + ...$ is the (supersymmetric) mass due to open string stretching between the branes of a distance $y$, plus winding masses and Kaluza-Klein masses in other tori; the full expression is
\bea
M_0^2 &\equiv& \sum_{j=1,2} M_0^2 (j) \nn\\
&=&\sum_{j=1,2} \bigg| \frac{n_j}{\V_j} +i \bigg( \frac{m_j T_2^j}{\ap \V_j} +\frac{ y_j}{2\pi \ap }\bigg) \bigg|^2 
\label{M0DEF}\eea
where $n_j, m_j$ are respectively the Kaluza-Klein and winding numbers in the $j^{th}$ torus.

\begin{table}[h]
\begin{center}
\begin{tabular}{|c|c|c|}\hline
Rep & $(\mathbf{1},-\mathbf{1})$ & $(-\mathbf{1},\mathbf{1})$ \\\hline 
Vector & $\epsilon$ & $\epsilon$ \\
LH Gaugino & $2\epsilon$ & $0$ \\
$\Phi_{1,2}$ & $\epsilon$ & $\epsilon$ \\
$\Phi_3$ & $3\epsilon$ & $-\epsilon$ \\
$\ov{\Phi}_3$ & $-\epsilon$ & $3\epsilon$ \\
$\Psi_{1,2}$ & $0$ & $2\epsilon$ \\
$\Psi_3$ & $2\epsilon$ & $0$
\\\hline\end{tabular}
\caption{$\ap \C{M}^2$ for $N \approx 4$ sectors.}\label{Napprox4spectrum}\end{center}
\end{table}

We can then calculate the Coleman-Weinberg effective potential. To do this we note \cite{Bachas:1995ik} $\mathrm{Str} \C{M}^{2n} =0 $ for $n < 4$, but we expect $\C{O}(1/|\epsilon|)$ levels below the string scale, or $\C{O}(\ap M^2/2|\epsilon|)$ below a cutoff scale $M^2$, so the potential should be $\C{O}(\epsilon^4)$. This follows from: 
\bea
64\pi^2 V &=& |I_{ab}| \sum_{j=1,2} \sum_{n_j, m_j}\sum_n \mathrm{Str} \C{M}_n^4 \log \C{M}^2_n \nn\\
&=& |I_{ab}|\sum_{j=1,2} \sum_{n_j, m_j}\sum_n \mathrm{Str} \bigg(- \frac{\C{M}^8}{12(M_0^2 + 2n|\epsilon|/\ap)^2}\bigg) + ... \nn\\
&=& -4 |I_{ab}|\sum_{j=1,2} \sum_{n_j, m_j}\frac{\epsilon^4}{(\ap)^2} \frac{1}{4\epsilon^2} \zeta (2, \ap M_0^2/2|\epsilon|) \nn\\
&=& -2 |I_{ab}|\sum_{j=1,2} \sum_{n_j, m_j}  \frac{|\epsilon|^3}{(\ap)^3 M_0^2} + \C{O}(\epsilon^4).
\eea

Since this is always negative and diverges as $M_0^2 \rightarrow 0$ we can infer that the system will inevitably be unstable. 

\subsection{$N\approx 2$}

From the effective field theory perspective, the only $N\approx 2$ sector we can consider has one angle equal to zero, so that the branes are parallel in the $j^{th}$ torus. However, this is much simpler than the $N\approx 4$ case, as there is no tower of light states. Here the low energy theory consists of a non-chiral pair of superfields with a $D$-term induced on one $U(1)$ by the brane rotation; the scalar masses are split by $\pm \epsilon/\ap$ while the fermions have no supersymmetry breaking masses. We can thus determine the effective potential to be 
\begin{align}
32\pi^2 V ~=~&  |I_{ab}|\sum_{n_j, m_j} (M_0^2(j) + \epsilon/\ap)^2 \log [M_0^2(j) + \epsilon/\ap]
\nn\\ &\qquad
+ (M_0^2 (j)- \epsilon/\ap)^2 \log [M_0^2(j) - \epsilon/\ap] - 2 M_0^4(j) \log M_0^2 (j) \nn\\
~=~& |I_{ab}| \sum_{n_j, m_j} (3 + 2 \log M_0^2 (j))\left(\frac{\epsilon}{\ap}\right)^2 - \frac{1}{6M_0^4 (j)} \left(\frac{\epsilon}{\ap}\right)^4 +...
\label{FTR}
\end{align}

\subsection{Tadpoles and adjoint scalar masses}

In the supersymmetric case, supersymmetry determines the strength of the coupling between the adjoints and the messenger states. This allows us to use the above effective potential computation to determine the adjoint scalar masses for the adjoints in directions where the branes are parallel by taking derivatives. Labeling the three complex adjoints as $\Sigma^j$, one has: 
\begin{align}
M_0^2 (j)~=~&\bigg| \frac{n_j}{\V_j} +i \bigg( \frac{m_j T_2^j}{\ap \V_j} +\frac{ y_j}{2\pi \ap } \bigg) + \sqrt{2} g \Sigma^j \bigg|^2 .
\end{align}
Clearly, it is wise to consider separately the real and imaginary components; write $\Sigma^j = \frac{1}{\sqrt{2}} ( \Sigma_1^j +i \Sigma_2^j)$ so that 
\bea
M_0^2 (j)&=& (\frac{n_j}{\V_j} + g \Sigma^j_1)^2 + \bigg( \frac{m_j T_2^j}{\ap \V_j} +\frac{ y_j}{2\pi \ap } + g \Sigma^j_2\bigg)^2 \nn\\
\partial_1 M_0^2 (j)&=& 2g (\frac{n_j}{\V_j} + g \Sigma^j_1) \nn\\
\partial_1^2 M_0^2 (j)&=& 2g^2 \nn\\
\partial_2 M_0^2 (j)&=& 2g (\frac{m_j T_2^j}{\ap \V_j} +\frac{ y_j}{2\pi \ap } + g \Sigma^j_2) \nn\\
\partial_2^2 M_0^2 (j)&=& 2g^2 .
\eea
Then, we can obtain the derivatives of the potential at zero adjoint vevs.

\subsubsection{$N\approx 2$ Sectors}

The single derivatives of the potential give singlet tadpoles:
\begin{align}
\partial_1 V ~=~& \frac{2g\epsilon^2}{16\pi^2(\ap)^2} \sum_{n_j, m_j} \frac{n_j}{\V_j M_0^2} ~=~ 0 \nn\\
\partial_2 V ~=~& \frac{2g\epsilon^2}{16\pi^2(\ap)^2} \sum_{n_j, m_j} \frac{\frac{m_j T_2^j}{\ap \V_j} +\frac{ y_j}{2\pi \ap }}{ M_0^2} 
\end{align}
They also receive other contributions from closed string exchange. However, note that they obey the property 
\begin{equation}
\partial_2 V ( -y_j) = - \partial_2 V (y_j) ,
\end{equation}
and thus we can cancel these potentially dangerous contributions by arranging for the supersymmetry breaking brane to have an image brane at the same but opposite distance from the ``visible'' brane. This is indeed automatic in the presence of an orientifold. 

We can now calculate the mass terms by taking  second derivatives of the potential:
\begin{align}
\partial_1^2 V ~=~& \frac{2g^2\epsilon^2}{16\pi^2(\ap)^2} \sum_{n_j, m_j} \frac{1}{M_0^4} \bigg[  \bigg( \frac{m_j T_2^j}{\ap \V_j} +\frac{ y_j}{2\pi \ap }\bigg)^2 - \bigg(\frac{n_j}{\V_j}\bigg)^2 \bigg]\nn\\
~\equiv~&  \frac{2g^2\epsilon^2}{16\pi^2(\ap)^2} X_{IR}^{N\approx 2} \nn\\
\partial_2^2 V ~=~& - \partial_1^2 V ~=~ - \frac{2g^2\epsilon^2}{16\pi^2 \ap} X_{IR}^{N\approx 2}\nn\\
\partial_1 \partial_2 V ~=~& 0.
\end{align}
We see that the field theory contributions from $N=2$ sectors inevitably lead to a tachyon, since there are  two states of opposite squared-masses $\pm \frac{2g^2\epsilon^2}{16\pi^2(\ap)^2} X_{IR}^{N\approx 2}$.

\subsubsection{$N\approx 4$ Sectors}

In this case we can take derivatives with respect to four real adjoints. We have define for simplicity $V^{(i,j,k,l)} \equiv \partial^i_{\Sigma^1_1} \partial^j_{\Sigma^1_2}\partial^k_{\Sigma^2_1}\partial^l_{\Sigma^2_2} V$ and recalling that 
$V= -\frac{|\epsilon|^3}{32\pi^2 (\ap)^3 M_0^2} + ... =-\frac{|\epsilon|^3}{32\pi^2 (\ap)^3 (M_0^2 (1) + M_0^2 (2))} + ...$, 
we have for the tadpoles:
\begin{align}
V^{(1,0,0,0)} ~=~&  0 \nn\\
V^{(0,1,0,0)} ~=~& |I_{ab}|\frac{|\epsilon|^3}{32\pi^2 (\ap)^3 }\sum_{n,m} \frac{2g}{M_0^4} \bigg[ \frac{m_1 T_2^1}{\ap \V_1} +\frac{ y_1}{2\pi \ap } \bigg] \nn\\
V^{(0,0,1,0)} ~=~&  0 \nn\\
V^{(0,0,0,1)} ~=~& |I_{ab}|\frac{|\epsilon|^3}{32\pi^2 (\ap)^3 }\sum_{n,m} \frac{2g}{M_0^4} \bigg[ \frac{m_2 T_2^2}{\ap \V_2} +\frac{ y_2}{2\pi \ap } \bigg].
\end{align}
These obey the same property as the $N\approx 2$ sectors of changing sign upon reflection of $y_j$, and thus in the presence of an orientifold we expect them to cancel in the same way. 

For the mass terms, we obtain:
\begin{align}
V^{(2,0,0,0)} ~=~& \frac{|\epsilon|^3 |I_{ab}|}{32\pi^2 (\ap)^3 }\sum_{n,m} \frac{2g^2}{M_0^6} \bigg[ -4 (\frac{n_1}{\V_1})^2 + M_0^2 \bigg] \nn\\
V^{(1,1,0,0)} ~=~& 0 \nn\\
V^{(1,0,1,0)} ~=~& 0 \nn\\
V^{(1,0,0,1)} ~=~& 0\nn\\
V^{(0,2,0,0)} ~=~&  \frac{|\epsilon|^3 |I_{ab}| }{32\pi^2 (\ap)^3 }\sum_{n,m} \frac{2g^2}{M_0^6} \bigg[ -4 (\frac{m_1 T_2^1}{\ap \V_1} +\frac{ y_1}{2\pi \ap })^2 + M_0^2 \bigg] \nn\\
V^{(0,1,1,0)} ~=~& 0 \nn\\
V^{(0,1,0,1)} ~=~& - \frac{|\epsilon|^3 |I_{ab}|}{32\pi^2 (\ap)^3 }\sum_{n,m} \frac{4g^2}{M_0^6} \bigg(\frac{m_1 T_2^1}{\ap \V_1} +\frac{ y_1}{2\pi \ap }\bigg) \bigg( \frac{m_2 T_2^2}{\ap \V_2} +\frac{ y_2}{2\pi \ap }\bigg) \nn\\
V^{(0,0,2,0)} ~=~&  \frac{|\epsilon|^3 |I_{ab}|}{32\pi^2 (\ap)^3 }\sum_{n,m} \frac{2g^2}{M_0^6} \bigg[ -4 (\frac{n_2}{\V_2})^2 + M_0^2 \bigg]\nn\\
V^{(0,0,1,1)} ~=~& 0 \nn\\
V^{(0,0,0,2)} ~=~&  \frac{|\epsilon|^3 |I_{ab}|}{32\pi^2 (\ap)^3 }\sum_{n,m} \frac{2g^2}{M_0^6} \bigg[ -4 (\frac{m_2 T_2^2}{\ap \V_2} +\frac{ y_2}{2\pi \ap })^2 + M_0^2 \bigg].
\end{align}
Let us define 
\begin{align}
A_{1,1} ~\equiv~ & (\frac{n_1}{\V_1}) \nn\\
A_{1,2} ~\equiv~ &(\frac{m_1 T_2^1}{\ap \V_1} +\frac{ y_1}{2\pi \ap }) \nn\\
A_{2,1} ~\equiv~ & (\frac{n_2}{\V_2}) \nn\\
A_{2,2} ~\equiv~ &(\frac{m_2 T_2^2}{\ap \V_2} +\frac{ y_2}{2\pi \ap }),
\end{align}
then we have a mass matrix 
\begin{equation}
\C{M}^2_{IR} ~\equiv~ \frac{|\epsilon|^3 g^2 |I_{ab}| }{32\pi^2 \ap}X_{IR}^{N\approx 4}
\end{equation}
where
{\small\begin{align}
X_{IR}^{N\approx 4}~ \equiv &\sum_{n_1,m_1,n_2,m_2} \frac{2}{M_0^6 (\ap)^2}\times\nonumber \\[3pt] 
&\hskip -1.3cm 
\left(\!\!\!\! \begin{array}{cccc} 
A_{1,2}^2 + A_{2,1}^2 + A_{2,2}^2 &0 & 0 & 0 \\[3pt] - 3 A_{1,1}^2\\[3pt]
0 & A_{1,1}^2 + A_{2,1}^2 + A_{2,2}^2 &0 & - A_{1,2} A_{2,2} \\[3pt] &- 3 A_{1,2}^2\\[3pt]
0 & 0 & A_{1,1}^2 + A_{1,2}^2 + A_{2,2}^2 &0 \\[3pt] & & - 3 A_{2,1}^2 \\[3pt]
0 & - A_{1,2} A_{2,2} & 0 & A_{1,1}^2 + A_{1,2}^2 + A_{2,1}^2\\[3pt] 
& & &- 3 A_{2,2}^2 \end{array} \!\!\!\!\right)
\label{Neq4parallelmasses}\end{align}}

The above sums  are dominated by their zero modes, so that we have non-negative squared-masses for $\Sigma_1^1$ and $\Sigma_1^2$. However, since the matrix has zero trace, there must be at least one negative eigenvalue if the mass-matrix is non-trivial. Since we require $y \ne 0$ to avoid tachyonic messengers, this will generically be the case.

%%%%%%%%%%%%%%%%%%%%%%%%%%%%%%%%%%%%%%%%%%%%%%%%%

\section{String CFT basics for intersecting branes}
\label{SEC:CFT}

To start our string computations, we require some background material. In all sections except \ref{SEC:SUGRA} we shall take the metric to be $\eta = (-1, 1, 1,...)$.
Throughout we shall take the annulus world-sheet to be $[0,1/2] \times [0,it/2]$.
For a given complex direction $X = \frac{1}{\sqrt{2}}(X_1 + i X_2)$ let us align one brane along the direction $X_1$. Then we must satisfy Neumann boundary conditions along $X_1$ ($\partial_\sigma X_1 = 0$) and Dirichlet boundary on $X_2$ ($\partial_\tau X_2 = 0$). For $w = \sigma + i\tau$, this corresponds to
\begin{align}
(\partial + \bar{\partial}) X_1 ~=~& 0 \nonumber \\
(\partial - \bar{\partial}) X_2 ~=~& 0
\end{align}
which can be rewritten
\begin{align}
\partial X + \bar{\partial} \ov{X} ~=~& 0 \nonumber \\
\partial \ov{X} + \bar{\partial} X ~=~& 0.
\end{align}

The above is valid for both boundaries if the second brane is parallel to the first. However, suppose instead that we have tilted the branes at an angle, so that we have $\partial_\sigma (\cos \pi \theta X_1 + \sin \pi \theta X_2) = 0 = \partial_\tau (-\sin \pi \theta X_1 + \cos \pi \theta X_2)$. Then, we have 
\begin{align}
e^{-\pi i\theta}\partial X + e^{\pi i\theta}\bar{\partial} \ov{X} ~=~& 0 \nonumber \\
e^{\pi i\theta}\partial \ov{X} + e^{-\pi i\theta}\bar{\partial} X ~=~& 0.
\end{align}
If this is at the boundary $\mathrm{Re}(w)=1/2$, we can use the doubling trick
\beq
\partial X = \left\{ \begin{array}{cc} \partial X(w), & \mathrm{Re}(w) > 0 \\ -\bar{\partial} \ov{X}(w), & \mathrm{Re}(w) < 0 \end{array} \right.
\eeq
and
\beq
\partial \ov{X} = \left\{ \begin{array}{cc} \partial \ov{X}(w), & \mathrm{Re}(w) > 0 \\ -\bar{\partial} X(w), & \mathrm{Re}(w) < 0 \end{array} \right.
\eeq
to obtain
\begin{align}
\partial X(w) ~&=~ e^{2\pi i \theta} X(w-1) \nonumber \\
\partial \ov{X}(w) ~&=~ e^{-2\pi i \theta} \ov{X}(w-1).
\end{align}

\subsection{Partition Functions}

Here we present the partition functions that we will need. The non-compact dimensions, together with the super-reparametrization ghosts, contribute in the spin-structure $\nu$:
\begin{align}
Z^{4d}_\nu ~=~& \frac{1}{(4\pi^2 \ap t)^2} \frac{\vartheta_\nu (0)}{\eta^3 (it/2)} .
\end{align}
In one compact complex dimension $j$ where the two branes are parallel, the partition function is
\begin{align}
Z_\nu^{\theta^j_{ab} =0} ~=~& \frac{\vartheta_\nu (0)}{\eta^3 (it/2)} Z_{cl}^j
\end{align}
where the classical piece is given by
\begin{align}
Z_{cl}^j ~=~ \sum_{n_j,m_j} e^{-S_{cl}^j} ~=~&\sum_{n_j,m_j} \exp\bigg[ - \frac{4\pi^3 \ap t}{L_j^2} | n_j + i\frac{T_2^j m_j}{\ap} + \frac{i y_j L_j}{4\pi^2\ap}|^2 \bigg],
\label{DefineZj}\end{align} 
with $y_j$ the separation distance of the branes in the perpendicular direction. 

When the branes are not parallel, the partition function is
\begin{align}
Z_\nu^{\theta^j_{ab} \ne0} ~=~& iI_{ab}^j \frac{\vartheta_\nu (\theta_{ab}^j it/2)}{\vt (\theta_{ab}^j it/2)}
\end{align}
where $I_{ab}^j$ is the number of intersections between the branes in that torus. 

The total partition function is given by 
\begin{align}
\frac{1}{2} \sum_{\nu=1}^4 \delta_\nu Z^{4d}_\nu \prod_{\kappa=1}^3 Z_\nu^\kappa ~\equiv~ \frac{1}{(4\pi^2 \ap t)^2} \frac{1}{2} \sum_{\nu} \delta_\nu \frac{\vartheta_\nu (0)}{\eta^3 (it/2)} \prod_{\kappa=1}^3 \vartheta_\nu (\theta_{ab}^\kappa it/2) .
\label{DEFINEZ}\end{align}
where $\delta_\nu = \{1,-1,1,-1\}$.

\subsection{Basic correlators for parallel branes}

Let us consider first parallel branes, where there is a zero mode. In turn, we must treat compact and non-compact dimensions separately.

\subsubsection{Non-compact dimensions}

Here we shall simply give general correlators for non-compact dimensions with Neumann or Dirichlet boundary conditions. These can be obtained from the standard expression on the covering torus via the doubling trick: 
\begin{align}
\langle X_i(z) X_i(w) \rangle_{\mr{{\cal A}}} ~=~  \half \bigg[& \langle X_i(z) X_i(w) \rangle_{\mr{{\cal T}}} \, \pm \, \langle X_i(1-\bar{z}) X_i(w) \rangle_{\mr{{\cal T}}} \nn\\
&\pm \langle X_i(z) X_i(1 - \bar{w})\rangle_{\mr{{\cal T}}} + \langle X_i(1-\bar{z}) X_i(1-\bar{w}) \rangle_{\mr{{\cal T}}} \bigg],
\end{align}
where the upper (lower) sign is for Neumann (Dirichlet) boundary conditions, and the subscripts $\cal A$, $\cal T$ denote the world-sheets annulus and torus, respectively. Let us say that $X_1$ obeys Neumann boundary conditions, and $X_2$ Dirichlet. Then, the corresponding non-vanishing correlators involve tangential or normal derivatives: $\partial_\tau X_i \leftrightarrow \dot{X}_i \equiv (\partial - \ov{\partial}) X_i, \partial_n X_i \equiv (\partial + \ov{\partial}) X_i$. In terms of elliptic theta functions
\begin{align}
\vt (z) ~\equiv~& \vartheta_{11} (z, \frac{it}{2}) \nn\\
~\equiv~& 2\sum_{n=0}^\infty (-1)^n e^{(n-1/2)^2 \pi t/2} \sin (2n + 1) \pi iz \nn\\
~=~& 2 e^{-\pi t/8} \sin \pi i z \prod_{m=1}^\infty (1-e^{-\pi m t})(1- e^{-2\pi z} e^{-\pi m t})(1-e^{2\pi z}e^{-\pi m t})
\end{align}
we have
\begin{align}
\bra \partial_\tau X_1 \partial_\tau X_1 \ket|_{z = - \ov{z}} ~=~& - 2\ap \partial_z \partial_w \log \vt (z - w) + \frac{8\pi \ap}{t} \nn\\
\bra \partial_n X_2 \partial_n X_2 \ket|_{z = - \ov{z}} ~=~& - 2\ap \partial_z \partial_w \log \vt (z - w)
\end{align}
Note that to restore the metric these should be multiplied by $\eta_{ij}$; for spacelike dimensions since we are taking  $\eta = (-1, 1, 1,...)$ this will always be one.

These often appear integrated over $z$. Observing that 
\begin{align}
\frac{\vt^\prime (z-w + it/2)}{\vt (z-w+it/2)} - \frac{\vt^\prime (z-w )}{\vt (z-w)} ~=~& - 2\pi i 
\end{align}
we have 
\begin{align}
\int_0^{it/2} dz - \frac{\ap}{2} \partial_z \partial_w \log \vt (z - w) ~=~& \bigg[ \frac{\ap}{2} \frac{\vt^\prime (z-w)}{\vt (z-w)} \bigg]^{it/2}_0 \nn\\
~=~& - \ap\pi i \nn\\
~=~& - \frac{2\pi\ap}{t} \frac{it}{2}.
\end{align}
Hence, we see 
\begin{align}
\int_0^{it/2} dz \bra \partial_\tau X_1 \partial_\tau X_1 \ket ~=~& 0\nn\\
\int_0^{it/2} dz \bra \partial_n X_2 \partial_n X_2 \ket ~=~& - 4\pi \ap i .
\end{align}

\subsubsection{Compact dimensions}

We will require the correlators for \emph{compact} dimensions, and therefore the zero modes on the torus may only take specific values. These are given by the classical part of the amplitude;  we split $X = X_{cl} + X_{qu}$ and note that, since $\bra X_{qu} \ket = 0$, there are no mixed correlators and we have separate ``quantum'' and ``classical'' correlators $\bra X^i X^i \ket_{cl} + \bra X^i X^i \ket_{qu}$. Thus for the quantum amplitude the zero mode should be excluded even in the Neumann directions, and we can write
\begin{align}
\bra \partial_\tau X_1 \partial_\tau X_1 \ket_{qu}|_{z = - \ov{z}} ~=~& - 2\ap \partial_z \partial_w \log \vt (z - w) \nn\\
\bra \partial_n X_2 \partial_n X_2 \ket_{qu}|_{z = - \ov{z}} ~=~& - 2\ap \partial_z \partial_w \log \vt (z - w) \nn\\
\langle \partial X (z) \partial \ov{X}(w) \rangle_{qu} ~=~& -\frac{\alpha'}{2} \partial_z\partial_w \log \theta_1 (z-w) \nn\\
\langle \partial X (z) \partial X(w) \rangle_{qu} ~=~& 0
\end{align}
and thus
\begin{align}
\int_0^{it/2} dz \bra \partial_\tau X_1 \partial_\tau X_1 \ket_{qu} ~=~& - 4\pi \ap i\nn\\
\int_0^{it/2} dz \bra \partial_n X_2 \partial_n X_2 \ket_{qu} ~=~& - 4\pi \ap i\nn\\
\int_0^{it/2} dz\bra \partial X \partial \ov{X}\ket_{qu} ~=~& -i\pi \ap\nn\\
\int_0^{it/2} dz\bra \partial X \partial X\ket_{qu} ~=~& 0.
\end{align}

Now the classical pieces (for complex coordinates on the torus $j$) are given by 
\begin{align}
\partial X^j ~=~&  \frac{1}{\sqrt{2}} 4\pi^2 \bigg[ n_j \frac{\ap}{L_j} + i \bigg( m_j \frac{T_2^j}{L_j} + \frac{y}{4\pi^2}\bigg) \bigg] \nn\\
~=~& \frac{1}{\sqrt{2}} 2\pi \bigg[ n_j \frac{\ap}{\V_j} + i \bigg( m_j \frac{T_2^j}{\V_j} + \frac{y}{2\pi}\bigg) \bigg] \nn\\
\ov{\partial} X^j ~=~&  \frac{1}{\sqrt{2}} 2\pi \bigg[ - n_j \frac{\ap}{\V_j} + i \bigg( m_j \frac{T_2^j}{\V_j} + \frac{y}{2\pi}\bigg) \bigg] \nn\\
\ov{\partial} \ov{X}^j ~=~& - \partial X \nn\\
\partial \ov{X}^j ~=~& - \ov{\partial} X,
\end{align}
where the wrapping/Kaluza-Klein numbers $n_j, m_j$ \emph{are those appearing in the classical action} $Z_{cl}^j$. 
Hence we can write, since $\partial X = \partial X_{qu} + \partial X_{cl}$ and $\bra \partial X_{qu} \ket = 0$
\begin{align}
\label{240}
\int_0^{it/2} dz \bra \partial X^j \partial \ov{X}^j\ket ~=~&\int_0^{it/2} dz \bra \partial X^j \partial \ov{X}^j\ket_{qu} + \bra \partial X^j \partial \ov{X}^j \ket_{cl} \nn\\
~=~&\sum_{n_j,m_j} \bigg[ -i\pi \ap + \pi^2 it \bigg(  n_j^2 \frac{(\ap)^2}{\V_j^2} + \bigg( m_j \frac{T_2^j}{\V_j} + \frac{y}{2\pi}\bigg)^2 \bigg) \bigg] \nn\\
&~~\times \exp \bigg[ - \frac{\pi \ap t}{\V_j^2} \bigg| n_j  + i \bigg( m_j \frac{T_2^j}{\ap} + \frac{y \V_j}{2\pi \ap}\bigg) \bigg|^2 \bigg],
\end{align}
and
\begin{align}
\int_0^{it/2} dz \bra \partial X^j \partial X^j\ket ~=~&\int_0^{it/2} dz \bra \partial X^j \partial X^j\ket_{qu} + \bra \partial X^j \partial X^j \ket_{cl} \nn\\
~=~&\sum_{n_j,m_j} \pi^2 it  \bigg[n_j \frac{\ap}{L_j} + i \bigg( m_j \frac{T_2^j}{L_j} + \frac{y}{4\pi^2}\bigg)\bigg]^2 e^{-S^j_{cl}}  \nn\\
~=~&\sum_{n_j,m_j} \pi^2 it \bigg(  n_j^2 \frac{(\ap)^2}{\V_j^2} - \bigg( m_j \frac{T_2^j}{\V_j} + \frac{y}{2\pi}\bigg)^2 \bigg) \bigg] e^{-S^j_{cl} }.
\end{align}
where the classical action is $S^j_{cl}=-\frac{t}{2\pi\alpha' }\partial X^j\ov\partial X^j$, given in the exponent of eq.~(\ref{240})

\subsection{Basic correlators for non-parallel branes}

For non-parallel branes, there is no zero mode and the correlators of the derivatives are just equal to those on the covering torus. The Green functions can then be determined similarly to that for orbifolds given in \cite{Poppitz:1998dj,Bain:2000fb,Benakli:2008ub,Conlon:2010qy}; they must satisfy:
\begin{align}
G^{\mr{\cal T}}_{\theta}\left(z-w+\tau\right)~ =~& G^{\mr{\cal T}}_{\theta}\left(z-w\right) \;, \nn \\
G^{\mr{\cal T}}_{\theta}\left(z-w+1\right) ~=~& e^{2\pi i \theta} G^{\mr{\cal T}}_{\theta}\left(z-w\right) \;
\end{align}
and
\beq
\underset{{z\rightarrow w}}{\mr{lim}}\;G^{\mr{\cal T}}_{\theta}\left(z-w\right)~ \sim~ -\frac{\alpha^\prime/2}{\left(z-w\right)^2} -\alpha^\prime Z^{-1}_{tw} \bra T(0) \ket \;. 
\eeq
To construct them, note that for $f(z) \equiv e^{2\pi i \theta z} \vartheta_1 (z + \theta it/2)$, 
\begin{align}
f(z+1) ~=~& - e^{2\pi i \theta} f(z) \nn\\
f(z + it/2) ~=~& - e^{\pi t/2} e^{-2\pi i z} f(z)
\end{align}
and consider the function:
\beq
G^{\mr{\cal T}}_{\theta}\left(z-w\right) = \frac{\ap}{2} \partial_z \bigg[ \frac{e^{2\pi i \theta (z-w)} \vartheta_1 (z - w + \theta it/2)}{\vartheta_1 (\theta i t/2)} \frac{\vartheta_1^\prime (0)}{\vartheta_1 (z-w)}\bigg] \;.
\eeq
Clearly it has the correct periodicity, and expanding around $(z-w) \sim 0$ we find
\begin{align}
G^{\mr{\cal T}}_{\theta}\left(z-w\right) \sim& -\frac{\alpha^\prime/2}{\left(z-w\right)^2} \nn\\
&~~+\frac{\ap}{2} \bigg(- 2\pi^2 \theta^2 + \frac{1}{2} \frac{\vt^{\prime\prime} (\theta i t/2)}{\vt(\theta it/2)} - \frac{1}{6} \frac{\vt^{\prime\prime\prime} (0)}{\vt^\prime(0)} +2\pi i \theta \frac{\vt^{\prime} (\theta i t/2)}{\vt(\theta it/2)}\bigg).
\end{align}
Comparing it to the twisted partition function
\begin{align}
\partial_t \log Z_{tw} ~\equiv~& \partial_t \log \bigg[ \frac{\exp[\pi \theta^2 t/2] \eta (it/2)}{\vt(\theta it/2)}\bigg] \nn\\
~=~& \partial_t \bigg\{\pi \theta^2 t/2 + \log \bigg[ \frac{ \theta_1^\prime (0)^{1/3}}{\vt( \theta i t/2)}\bigg] \bigg\} \nn\\
~=~& \frac{1}{2}\pi \theta^2 - \theta i/2 \frac{\vt^\prime(\theta i t/2)}{\vt(\theta it/2)} + \frac{1}{8\pi} \bigg[\frac{1}{3} \frac{\theta_1^{\prime\prime\prime}(0)}{\theta_1^\prime(0)} - \frac{\theta_1^{\prime\prime}(\theta i t/2)}{\theta_1(\theta it/2)} \bigg] \nn\\
~=~& -\frac{1}{4\pi} \bigg[-2\pi^2 \theta^2 + 2\pi i\theta \frac{\vt^\prime(\theta i t/2)}{\vt(\theta it/2)} - \frac{1}{6} \frac{\theta_1^{\prime\prime\prime}(0)}{\theta_1^\prime(0)} +\frac{1}{2}\frac{\theta_1^{\prime\prime}(\theta it/2)}{\theta_1(\theta it/2)}\bigg] 
\end{align}
and noting that $\re(Z^{-1}_{tw}  \bra T(0)\ket) = 2\pi \partial_t \log Z_{tw}$ we find complete agreement. Hence we have
\begin{align}
\langle \partial X (z) \partial \ov{X}(w) \rangle ~=~& Z_{tw}  G^{\mr{\cal T}}_{\theta}\left(z-w\right). \nn\\
\langle \partial X (z) \partial X(w) \rangle ~=~& 0.
\end{align}
Crucially, the first term is a derivative of a periodic function on the boundary of the annulus, and is vanishing upon integration: 
\beq
\int_0^{it/2} dz \bra \partial X(z) \partial \ov{X}(0) \ket ~=~ 0.
\label{GintIsZero}\eeq

\subsection{Vertex Operators}

The vertex operator associated to a scalar in the adjoint representation reads
\begin{align}
V^0_{X^i} ~=~& 2g \bigg( \ap (k\cdot \psi )\Psi^i + i \partial X^i\bigg),
\end{align}
and corresponds to the gauge boson vertex normalization
\begin{align}
V_A^0 ~=~& g \bigg( - \dot{X}^\mu + 2 \ap i (k\cdot \psi )\psi^\mu \bigg)
\end{align}
where we neglect Chan-Paton factors.

\section{Effective potential}
\label{SEC:POTENTIAL}

For adjoint scalars associated with moduli Wilson lines, the one-loop induced  mass can be extracted from the  effective potential. The string vacuum amplitude is given by
\begin{align}
\A_0 ~=~ \bra 1 \ket ~=~& i \int_0^\infty \frac{dt}{2t} \frac{1}{(4\pi^2 \ap t)^2} \tr^\prime (\exp [-\pi t L_0]) 
\end{align}
which gives the effective potential via $V = i \A_0$; it matches the Coleman-Weinberg result when we recall that $L_0 = \ap H_0 $ and substitute $\ap \pi t \rightarrow t$:
\begin{align}
V ~= & - \frac{1}{32 \pi^2} \int_0^\infty \frac{dt}{t^3} \tr^\prime (\exp [- t H_0]). 
\end{align}
However, for string computations we require the form
\begin{align}
V ~= &  - \frac{1}{32 \pi^2 (\ap)^2 } \int \frac{dt}{t} \sum_\nu \delta_\nu \frac{1}{2} Z_\nu (it/2)
\end{align}
where we have now included the sum over spin structures and the factor of $1/2$ from the GSO projection. To compute the contribution to the potential $V_{ab}$ for states stretched between two branes $a$ and $b$ we must also include both orientations of the string, which introduces a factor of two, giving
\begin{align}
V_{ab} =& - \frac{1}{16 \pi^4 (\ap)^2 } \int \frac{dt}{t^3} \sum_\nu \delta_\nu \frac{1}{2} Z(t) \theta_\nu (0) \prod_{\kappa=1}^3 \theta_\nu (\theta_{ab}^\kappa it/2) \nn\\
=&  - \frac{1}{16 \pi^4 (\ap)^2 } \int \frac{dt}{t^3} Z(t) \vt (\epsilon it) \prod_{\kappa=1}^3 \vt ((\theta_{ab}^\kappa - \epsilon)it/2)
\end{align}
where we have used $Z(t)$ as defined in equation (\ref{DEFINEZ}). 

As the simplest case, consider  $N\approx 2$ sectors where the branes are parallel in one, the $j^{th}$, torus.  Here, the  low energy gauge theory consists of a non-chiral pair of superfields charged under a $U(1)$ with non-vanishing $D$-term induced   by the brane rotation; the scalar masses for states localised at the brane intersections are then split by $\pm \epsilon/\ap$ while the fermions have no supersymmetry breaking masses. 

The string computation gives
\begin{align}
V ~=~& I_{ab}\int_0^\infty \frac{dt}{t} \frac{1}{(4\pi^2 \ap t)^2} \frac{\vt (\epsilon it/2)^2 \vt ((\theta + \epsilon)it/2) \vt ( (-\epsilon - \theta)it/2)}{\eta^6 (it/2) \vt ((\theta + 2\epsilon)it/2) \vt (( -\theta)it/2)} Z_{cl}^j \nn\\
~\rightarrow~& -|I_{ab}|\frac{\epsilon^2}{16\pi^2 (\ap)^2} \int_{\pi/\ap \Lambda^2}^\infty \frac{dt}{t} Z_{cl}^j \nn\\
&=~ -|I_{ab}|\frac{\epsilon^2}{16\pi^2 (\ap)^2} \int_{1/\pi\ap \Lambda^2}^\infty \frac{dt}{t} e^{- \pi t \ap M_0^2} \nn\\
&=~ \frac{\epsilon^2}{16\pi^2 (\ap)^2}|I_{ab}|  \log M_0^2/\Lambda^2 + ...
\end{align}
where $Z_{cl}^j$ is defined in (\ref{DefineZj}), $M_0$ is that defined in (\ref{M0DEF}), and we see that we obtain perfect agreement with the field theory result (\ref{FTR}).

%%%%%%%%%%%%%%%%%%%%%%%%%%%%%%%%%%%%%%%%%%%%%%%%%%%%%%%%%%%%%%%%
\section{Masses from string two-point amplitudes}
\label{SEC:TWOPOINT}

For string amplitudes that are proportional to $k^2$, there are three ways that a mass term (finite when $k^2 \rightarrow 0$) can be generated. Firstly, and most commonly, is the closed string channel. As $t\rightarrow 0$ the amplitude becomes $\sim k^2 \int \frac{dt}{t^2} \chi(z)$; writing $t = 1/l$ this becomes $\sim k^2 \int dl \chi(z)$. This is the form found, for example, in generating masses for $U(1)$ gauge bosons (or adjoints) where the operators are on opposite boundaries \cite{akr}; in this case
\begin{align}
\chi(\frac{1}{2} + ixt/2) \;\xrightarrow{l\rightarrow \infty}\;& e^{-\pi \ap k^2 l}
\end{align}
and $\A = k^2 \int^\infty_a dl e^{-\pi \ap k^2 l} \rightarrow \frac{1}{\pi \ap}$. Such masses correspond to tree-level closed string exchange. However, since these contribute only to $U(1)$ gauge bosons and adjoint singlets we shall not be interested in these contributions. Rather, we shall consider the contributions where the vertex operators are on the same boundary. In this case, the above regulation of the amplitude is not possible; instead
\begin{equation}
\chi(ix/2l)\;\xrightarrow{l\rightarrow \infty}\; (2\sin \pi x)^{-2\ap k^2} 
\end{equation}
and so if there is a prefactor of $k^2$ masses are not generated in this way; instead we have a tadpole. The presence of such tadpoles indicates a false vacuum; they can either be removed by calculating in the true vacuum, or in principle by summing all contributions in the false vacuum \cite{Tadpoles}. We shall simply keep track of them by defining
\begin{equation}
K ~\equiv~ \pi \ap k^2 \int_0^\infty dl
\label{DefineK}\end{equation}
as the coefficient of these, so that the amplitude can be written
\begin{align}
\mathcal{A} ~\supset&~ -i A_{UV} K + ... 
\end{align}
Note that we could regulate such amplitudes by including a mass $M$ for the closed string states; then we would write
\begin{align}
K ~&\rightarrow~ \pi \ap k^2 \int_0^\infty dl e^{-\pi \ap M^2 l} \nn\\
~&~~~\rightarrow~ \frac{k^2}{M^2}.
\end{align}

A second source of masses can occur as $t \rightarrow \infty$ if the amplitude behaves as $\A \sim k^2 \int dt \chi(z)$; this corresponds to massless states in the loop, and is somewhat uncommon, although it was found in \cite{Conlon:2010xb}. 

Finally we can have world-sheet poles. Single poles give us momentum poles via
\beq
\int d(z_1 - z_2) \left( \frac{\vartheta_1(z_1 - z_2)}{\vartheta_1^{'}(0)} \right)^{-1 - 2 \ap k^2}
\sim \int d(z_1 - z_2) (z_1 - z_2)^{-1 - 2 \ap k^2} \to \frac{1}{2\ap k^2}
\eeq 
whereas double poles do not contribute as $k^2 \rightarrow 0$ due to analytic continuation in $k^2$. Our amplitudes will superficially appear to have both double and single poles. However, there may be poles both at $z_1 = z_2$ and $z_1 = it/2 + z_2$, and in principle they could cancel. We can write our amplitudes as
\beq
\A = \int dt g(t) \int_0^{it/2} dz f(z) = g(t) \int_0^{it/4} dz \left[ f(z) + f(it/2 - z)\right]
\eeq
and the single poles may cancel between the two contributions. In fact, all of our amplitudes are periodic in $z \rightarrow z + it/2$, giving rise to
\beq
\A = \int dt g(t) \int_0^{it/4} dz \left[ f(z) + f( - z)\right].
\eeq
Below, we will find that
\beq
f(z) = \chi(z) e^{4\pi i \epsilon z} \frac{\vt (z + \epsilon it/2)^2}{\vt(z)^2}
\eeq
which will be the generic case for our non-supersymmetric amplitudes. Moreover, we will be able to write
\beq
f(z) \equiv \chi(z)\frac{h(z)\vt^\prime(0)^2}{\vt(z)^2}.
\eeq
Using the fact that $\chi(z)$ is even, we see that 
\begin{align}
f(z) + f(-z) \sim& \chi(z)^{k^2}\frac{2 h(0)}{z^2} + \C{O}(1)
\end{align}
and so there is a double pole, which gives vanishing contribution by the usual left-right conformal regularization, but no single pole. Therefore there are no world-sheet poles in our amplitudes, apart from the UV ones. Note that the above reasoning would break down for non-periodic amplitudes.

%%%%%%%%%%%%%%%%%%%%%%%%%%%%%%%%%%%%%%%%%%%%%%%%%%%%%%

\section{Stringy contributions to  adjoint scalar masses?}
\label{SEC:STRINGY}

Here we would like to see if there can be any specifically stringy contributions to adjoint scalar masses, that cannot be reproduced from the Coleman-Weinberg potential. For this we need to calculate two 2-point amplitudes involving the scalars $\Sigma^i$:
\begin{align}
\A_{\Sigma^i \Sigma^j} =& -\frac{g^2}{2} \int \frac{dt}{t^2} \frac{1}{16\pi^4 (\ap)^2} \int_0^{it/2} dz \chi(z) \bigg[ 4 \bra \partial X^i (z) \partial X^j (0) \ket_{cl}  \bigg] \nn\\
\A_{\Sigma^i \ov{\Sigma}^{j\ne i }} =& -\frac{g^2}{2} \int \frac{dt}{t^2} \frac{1}{16\pi^4 (\ap)^2} \int_0^{it/2} dz \chi(z) \bigg[ 4 \bra \partial X^i (z) \partial \ov{X}^j (0) \ket_{cl}  \bigg] \nn\\
\A_{\Sigma^i \ov{\Sigma}^i} =& -\frac{g^2}{2} \int \frac{dt}{t^2} \frac{1}{16\pi^4 (\ap)^2} \int_0^{it/2} dz \chi(z)\bigg[ 4\bra \partial X^i (z) \partial \ov{X}^i (0) \ket  \nn\\
&\hskip 4.5cm - 4(\ap)^2 k^2 \bra \psi(z) \psi(0) \ket \bra \Psi^i (z) \ov{\Psi}^i (0)\ket \bigg].
\end{align}
In the first line there is no $\bra \Psi^i \Psi^j\ket$ contribution, nor quantum part to the $\bra \partial X^i \partial X^i \ket$ amplitude. Note that amplitudes  $\A_{\Sigma^i \Sigma^{j\ne i} }, \A_{\Sigma^i \ov{\Sigma}^{j\ne i }} $ only contribute because they have a classical part (the quantum part of the amplitudes is zero) which corresponds to a contribution that can be understood from the field theory; these shall be dealt with in section \ref{SEC:NONSTRINGY}. In this section we shall calculate the above amplitudes with $i = j$.

\subsection{The contribution from world-sheet fermions}

Let us first deal with the world-sheet-fermionic contribution:
\begin{align}
\A_{\Sigma^i \ov{\Sigma}^i}^\Psi ~\equiv~ & 2 g^2 k^2 \int \frac{dt}{t^2} \frac{1}{16\pi^4 } \int_0^{it/2} dz \chi(z)\bra \psi(z) \psi(0) \ket \bra \Psi^i (z) \ov{\Psi}^i (0)\ket \nn\\
~=~& 2 g^2 k^2 \int \frac{dt}{t^2} \frac{1}{16\pi^4 }\frac{ (\vt^\prime(0))^2 Z(t)}{\eta^3 (it/2)} \int_0^{it/2} dz \chi(z) e^{2\pi i \theta_{ab}^i z} \nn\\
~~&\times  \sum_{\nu \ne 1} \frac{\delta_\nu}{2} \vartheta_\nu (z) \vartheta_\nu (z + \theta_{ab}^i it/2) \vartheta_\nu (\theta_{ab}^j it/2) \vartheta_\nu (\theta_{ab}^k it/2) \nn\\
~\equiv~& \A_{\Sigma^i \ov{\Sigma}^i}^{\Psi_0} + \A_{\Sigma^i \ov{\Sigma}^i}^{\Psi_1}
\end{align}
where we have defined
\begin{align}
\A_{\Sigma^i \ov{\Sigma}^i}^{\Psi_0} ~\equiv~& 2 g^2 k^2\int \frac{dt}{t^2} \frac{1}{16\pi^4 } \frac{ (\vt^\prime(0))^2 Z(t)}{\eta^3 (it/2)}  \vt (  (\theta_{ab}^j -\epsilon)it/2) \vt((\theta_{ab}^k -\epsilon)it/2) \nn\\
&~\times\int_0^{it/2} dz \chi(z)  e^{2\pi i \theta_{ab}^i z} \frac{\vt (z + \epsilon it/2)\vt(z + (\theta_{ab}^i -\epsilon)it/2)}{\vt(z)^2} \nn\\
\A_{\Sigma^i \ov{\Sigma}^i}^{\Psi_1} ~\equiv~& - g^2 k^2\int \frac{dt}{t^2} \frac{1}{16\pi^4 } \frac{ (\vt^\prime(0))^2 Z(t)}{\eta^3 (it/2)}  \vt (  \theta_{ab}^j it/2) \vt(\theta_{ab}^k it/2) \nn\\
&~\times\int_0^{it/2} dz \chi(z)  e^{2\pi i \theta_{ab}^i z} \frac{\vt (z)\vt(z + \theta_{ab}^iit/2)}{\vt(z)^2}
\end{align}

For the different cases:
\begin{align}
Z_{N\approx 4} (t) ~=~& I_{ab} \bigg[ \eta^6 (it/2) (-i) \vt (\epsilon i t) \bigg]^{-1} \prod_{j\ne i} Z_{cl}^j \nn\\
Z_{N\approx 1} (t) ~=~&I_{ab} \bigg[ (-i)^3 \vt ( \theta_{ab}^i i t/2)\vt ( \theta_{ab}^j i t/2)\vt ( \theta_{ab}^k i t/2) \bigg]^{-1} \nn\\
Z_{N\approx 2}^{\theta_{ab}^j =0} (t) ~=~& I_{ab}\bigg[ \eta^3 (it/2) (-i)^2 \vt (\theta_{ab}^i i t/2) \vt (\theta_{ab}^k i t) \bigg]^{-1} Z_{cl}^j \nn\\  
Z_{N\approx 2}^{\theta_{ab}^j \ne 0} (t) ~=~& I_{ab}\bigg[ (-i)^3 \vt ( \theta_{ab}^i i t/2)\vt ( \epsilon i t)\vt ( \theta_{ab}^k i t/2) \bigg]^{-1}
\end{align}
where $Z_{cl}^j$ is the classical contribution defined in equation (\ref{DefineZj}).

As there are no world-sheet poles in the above amplitude, let us examine first the possible infrared singularities, as $t\rightarrow \infty$. Firstly we see that in the $N \approx 4$ and $N \approx 2$ with some $\theta_{ab}^j = 0$ these are impossible, as $Z \rightarrow e^{- \pi \ap M_0^2 t}$. The other two cases can be treated as follows:
\begin{align}
\A_{\Sigma^i \ov{\Sigma}^i}^{\Psi_0} ~=~& 2i g^2I_{ab}  k^2\int \frac{dt}{t} \frac{2}{16\pi^2 } \eta^3 (it/2) \frac{  \vt (  (\theta_{ab}^j -\epsilon)it/2) \vt((\theta_{ab}^k -\epsilon)it/2)}{ \vt (  (\theta_{ab}^j)it/2) \vt((\theta_{ab}^k)it/2)} \nn\\
&~~\times \int_0^{1} dx \chi(xit/2)  e^{-\pi  \theta_{ab}^i x t} \frac{\vt (x + \epsilon) it/2)\vt((x + \theta_{ab}^i -\epsilon)it/2)}{\vt (\theta_{ab}^i it/2) \vt(xit/2)^2}\nn\\
~&\rightarrow~ -2i g^2 k^2I_{ab}\int \frac{dt}{t} \frac{2}{16\pi^2 } \int_0^{1/2} dx   e^{-\pi  \theta_{ab}^i x t} \nn\\
&\rightarrow~ -2i g^2 k^2I_{ab}\int \frac{dt}{t^2} \frac{2}{16\pi^2 } \frac{1}{\pi \theta_{ab}^i} \nn\\
&\rightarrow~ 0.
\end{align}
Also
\begin{align}
\A_{\Sigma^i \ov{\Sigma}^i}^{\Psi_1} ~=~& - i g^2 k^2I_{ab}\int \frac{dt}{t} \frac{2}{16\pi^2 } \eta^3 (it/2) \frac{  \vt (  \theta_{ab}^jit/2) \vt(\theta_{ab}^k it/2)}{ \vt (  \theta_{ab}^jit/2) \vt(\theta_{ab}^kit/2)} \nn\\
&~~\times \int_0^{1} dx \chi(xit/2)  e^{-\pi  \theta_{ab}^i x t} \frac{\vt (x it/2)\vt((x + \theta_{ab}^i it/2)}{\vt (\theta_{ab}^i it/2) \vt(xit/2)^2}\nn\\
&\rightarrow~ -2i g^2 k^2I_{ab}\int \frac{dt}{t} \frac{2}{16\pi^2 } \int_0^{1/2} dx   e^{-\pi  \theta_{ab}^i x t} \nn\\
&\rightarrow~ -2i g^2 k^2I_{ab}\int \frac{dt}{t^2} \frac{2}{16\pi^2 } \frac{1}{\pi \theta_{ab}^i} \nn\\
&\rightarrow~ 0.
\end{align}
This corresponds to the fact that these adjoints have no renormalizable couplings to the corresponding light matter fields. 

Now let us consider the closed string poles, transforming to $t= 1/l$:
\begin{align}
\vt ( x it/2, it/2 ) ~=~& i (t/2)^{-1/2} \exp [ \frac{ \pi x^2 t}{2} ] \vt (x, 2il) \nn\\
~=~&i (2l)^{1/2}  \exp [ \frac{ \pi x^2 }{2l} ] \vt (x, 2il) \nn\\
\eta(it/2)^3 ~=~& (2l)^{3/2} \eta(2il)^3 \nn\\
&\rightarrow~ e^{-\pi l /2}\nn\\
\chi(xit/2) ~\rightarrow~&  (2\sin \pi x)^{-2\ap k^2}.
\end{align}

We can then write 
\begin{align}
\A_{\Sigma^i \ov{\Sigma}^i}^{\Psi_0} ~=~& -2i g^2 k^2 \int dl \frac{4}{16\pi^2 } \eta^3 (2il) \tilde{Z}(l)  \vt (  \theta_{ab}^j -\epsilon) \vt(\theta_{ab}^k -\epsilon) \nn\\
&~~~~\times\int_0^{1} dx \chi(x)  \frac{\vt (x + \epsilon)\vt(x + \theta_{ab}^i -\epsilon)}{\vt(x)^2} \nn\\
\A_{\Sigma^i \ov{\Sigma}^i}^{\Psi_1} ~=~& i g^2 k^2 \int dl \frac{4}{16\pi^2 } \eta^3 (2il) \tilde{Z}(l)  \vt (  \theta_{ab}^j) \vt(\theta_{ab}^k ) \nn\\
&~~~~\times\int_0^{1} dx \chi(x)  \frac{\vt (x )\vt(x + \theta_{ab}^i )}{\vt(x)^2},
\end{align}
where we have defined $\tilde{Z} (l) \equiv (2l)^{3/2} Z (t)$.
Now as $t\rightarrow 0$, 
\begin{align}
Z_{cl}^j ~=~& \frac{1}{4\pi^2 t} \frac{L_j^2}{T_2^j} \sum_{n_j,m_j} \exp\bigg[ - \frac{L_j^2}{4\pi \ap t} | n_j + i\frac{\ap m_j}{T_2^j}|^2\bigg] \exp [- im_j \frac{y_j L_j}{2\pi T_2} ] \nn\\
&\rightarrow~  2l \frac{\V_j^2}{2T_2^j}
\end{align}
and so 
\begin{align}
\tilde{Z}_{N\approx 4} (l) ~=~& I_{ab}^i \bigg[ \eta^6 (2il)  \vt (2\epsilon ) \bigg]^{-1} \prod_{j\ne i} (2l)^{-1} \tilde{Z}_{cl}^j \nn\\
&\rightarrow~ |I_{ab}^i| \bigg[ \eta^6 (2il)  \vt (2|\epsilon| ) \bigg]^{-1} \prod_{j\ne i} \frac{\V_j^2}{2T_2^j} \nn\\
\tilde{Z}_{N\approx 1} (l) ~=~& I_{ab} \bigg[  \vt ( \theta_{ab}^i )\vt ( \theta_{ab}^j )\vt ( \theta_{ab}^k ) \bigg]^{-1} \nn\\
\tilde{Z}_{N\approx 2}^{\theta_{ab}^j =0} (l) ~=~& \bigg[ \eta^3 (2il) \vt (\theta_{ab}^i ) \vt (\theta_{ab}^k ) \bigg]^{-1} (2l)^{-1} I_{ab}^j \tilde{Z}_{cl}^j \nn\\  
&\rightarrow~  \bigg[ \eta^3 (2il) \vt (\theta_{ab}^i ) \vt (\theta_{ab}^k ) \bigg]^{-1} I_{ab}^j \frac{\V_j^2}{2T_2^j} \nn\\
\tilde{Z}_{N\approx 2}^{\theta_{ab}^j \ne 0} (l) ~=~& I_{ab} \bigg[ \vt ( \theta_{ab}^i )\vt ( 2\epsilon )\vt ( \theta_{ab}^k ) \bigg]^{-1}
\end{align}

In order to determine the $UV$ tadpoles we require the limiting behaviour of the amplitude as $l \rightarrow \infty$; to this end we define
\beq
\tilde{Z}^\infty \equiv \lim_{l \rightarrow \infty} e^{-3\pi l/2} \tilde{Z} (l)
\eeq
to obtain
\begin{align}
\A_{\Sigma^i \ov{\Sigma}^i}^{\Psi_0} =& -2i g^2 k^2 \int dl \frac{4}{16\pi^2 } \tilde{Z}^\infty  4 \sin \pi (\theta_{ab}^j - \epsilon) \sin \pi (\theta_{ab}^k -\epsilon) \nn\\
&~~~\times\int_0^{1} dx   \frac{\sin \pi (x + \epsilon)\sin \pi (x + \theta_{ab}^i -\epsilon)}{\sin (\pi x)^2} \nn\\
=&  -2i g^2\frac{K}{\pi^3 \ap} \tilde{Z}^\infty \sin \pi (\theta_{ab}^j - \epsilon) \sin \pi (\theta_{ab}^k -\epsilon)\!\!\int_0^{1} \!\! dx   \frac{\sin \pi (x + \epsilon)\sin \pi (x + \theta_{ab}^i -\epsilon)}{\sin (\pi x)^2} \nn\\
&+ \C{O} (k^2) 
\end{align}
containing the closed string tapole contribution $K$ defined in equation (\ref{DefineK}). 
Then we have
\begin{align}
\int_0^1 dx &\frac{\sin \pi (x + \epsilon)\sin \pi (x + \theta_{ab}^i -\epsilon)}{\sin (\pi x)^2} \nn\\
~=~& 2\int_0^{1/2} dx\bigg[ \frac{\sin^2\pi x \cos \pi \epsilon \cos \pi (\theta_{ab}^i - \epsilon) + \cos^2\pi x \sin \pi \epsilon \sin \pi (\theta_{ab}^i - \epsilon)}{\sin^2 \pi x} \bigg] \nn\\
~=~& \cos \pi \epsilon \cos \pi (\theta_{ab}^i - \epsilon) -  \sin \pi \epsilon \sin \pi (\theta_{ab}^i - \epsilon) \nn\\
~=~& \cos \pi \theta_{ab}^i
\end{align}
and thus
\begin{align}
\A_{\Sigma^i \ov{\Sigma}^i}^{\Psi_0} =&  -i g^2 K \frac{4}{16\pi^3 \ap} (8\tilde{Z}^\infty) \sin \pi (\theta_{ab}^j - \epsilon) \sin \pi (\theta_{ab}^k -\epsilon) \cos \pi \theta_{ab}^i + \C{O}(k^2) \nn\\
\A_{\Sigma^i \ov{\Sigma}^i}^{\Psi_1} =&~~ i g^2 K \frac{2}{16\pi^3 \ap} (8\tilde{Z}^\infty) \sin \pi \theta_{ab}^j  \sin \pi \theta_{ab}^k  \cos \pi \theta_{ab}^i + \C{O}(k^2).
\end{align}

Now 
\begin{align}
8\tilde{Z}_{N\approx 4}^\infty ~=~& 8 I_{ab}^i  \bigg[ 2\sin (2\pi\epsilon ) \bigg]^{-1} \prod_{j\ne i} \frac{\V_j^2}{2T_2^j} \nn\\ 
~=~&|I_{ab}^i| \bigg[ |\sin (2\pi\epsilon )| \bigg]^{-1} \prod_{j\ne i} \frac{\V_j^2}{T_2^j} \nn\\ 
8\tilde{Z}_{N\approx 1}^\infty  ~=~& I_{ab} \bigg[ \sin (\pi \theta_{ab}^i )\sin ( \pi\theta_{ab}^j )\sin ( \pi \theta_{ab}^k ) \bigg]^{-1} \nn\\
8\tilde{Z}_{N\approx 2}^{\infty,\; \theta_{ab}^j =0} ~=~& \bigg[\sin (\pi \theta_{ab}^i ) \sin (\pi \theta_{ab}^k ) \bigg]^{-1} I_{ab}^i I_{ab}^k \frac{\V_j^2}{T_2^j}  \nn\\  
8\tilde{Z}_{N\approx 2}^{\infty,\; \theta_{ab}^j \ne 0}  ~=~& I_{ab} \bigg[ \sin ( \pi \theta_{ab}^i )\sin ( 2\pi \epsilon )\sin ( \pi\theta_{ab}^k ) \bigg]^{-1}
\end{align}
It is straightforward to show that for all of the cases
\begin{align}
8\tilde{Z}^\infty =& \prod_{\kappa=3}^3 \frac{\V_a^\kappa \V_b^\kappa}{T_2^\kappa}.
\end{align}

Hence
\begin{align}
\A_{\Sigma^i \ov{\Sigma}^i}^{N\approx 4} =& -i g^2 K \frac{4}{16\pi^3 \ap} I_{ab}^i \frac{\sin \pi (- \epsilon) \sin \pi ( -\epsilon) \cos 2\pi \epsilon}{\sin (2\pi\epsilon )} \prod_{j\ne i} \frac{\V_j^2}{T_2^j}\\
\A_{\Sigma^i \ov{\Sigma}^i}^{N\approx 2,\;\theta_{ab}^j =0} =& -i g^2 K \frac{4}{16\pi^3 \ap} \frac{\sin \pi ( - \epsilon) \sin \pi (\theta_{ab}^k -\epsilon) \cos \pi \theta_{ab}^i}{\sin (\pi \theta_{ab}^i ) \sin (\pi \theta_{ab}^k )} I_{ab}^i I_{ab}^k\frac{\V_j^2}{T_2^j} 
\end{align}
Most importantly for checking the normalization are the amplitudes that survive in the supersymmetric limit:
\begin{align}
\A_{\Sigma^i \ov{\Sigma}^i}^{N\approx 1} =& -i g^2 K \frac{2}{16\pi^3 \ap} I_{ab} \frac{\big( 2 \sin \pi (\theta_{ab}^j - \epsilon) \sin \pi (\theta_{ab}^k -\epsilon) - \sin \pi \theta_{ab}^j  \sin \pi \theta_{ab}^k \bigg)\cos \pi \theta_{ab}^i }{ \sin (\pi \theta_{ab}^i )\sin ( \pi\theta_{ab}^j )\sin ( \pi \theta_{ab}^k ) }
\end{align}
and
\begin{align}
\A_{\Sigma^i \ov{\Sigma}^i}^{N\approx 2,\;\theta_{ab}^j \ne0} =& -i g^2 K \frac{2}{16\pi^3 \ap} I_{ab} \times\nn\\
&~~~~\frac{\bigg( 2\sin \pi (\theta_{ab}^j - \epsilon) \sin \pi (\theta_{ab}^k -\epsilon) - \sin \pi \theta_{ab}^j  \sin \pi \theta_{ab}^k \bigg)\cos 2\pi \epsilon}{ \sin (2 \pi \epsilon )\sin ( \pi\theta_{ab}^j )\sin ( \pi \theta_{ab}^k ) }.
\end{align}

In summary the amplitudes involving the worldsheet fermion insertions can only contribute to tadpoles and not to masses for the adjoints.

\subsection{The bosonic contribution}

Let us now deal with the bosonic contribution, which we define
\begin{align}
\A_{\Sigma^i \ov{\Sigma}^i}^X \equiv&  -\frac{g^2}{2} \int \frac{dt}{t^2} \frac{1}{16\pi^4 (\ap)^2} \int_0^{it/2} dz \chi(z) 4 \bra \partial X^i (z) \partial \ov{X}^i (0) \ket \nn\\
=& -\frac{g^2}{2} \int \frac{dt}{t^2} \frac{1}{16\pi^4 (\ap)^2} Z(t)  \frac{\vt (\epsilon it)}{\eta^3 (it/2)} \prod_{\kappa=1}^3 \vt ((\theta_{ab}^\kappa - \epsilon)it/2)\int_0^{it/2} dz \chi(z) 4 G_{\theta_{ab}^i} (z).
\end{align}
Now, due to (\ref{GintIsZero}) after the integration over $z$ we will have something of order $k^2$ due to the presence of $\chi(z)$ in the integral. However, we may still obtain a tadpole from $t \rightarrow 0$. For this we need to transform to the closed string channel, noting that in our conventions $\theta_1^\prime (0) = 2\pi \eta^3 (it/2)$:
\begin{align}
G_{\theta_{ab}^i} (xit/2) ~=~& \frac{\ap}{2} \partial_z \bigg[ \frac{e^{-\pi  \theta x t} \vartheta_1 (xit/2 + \theta it/2)}{\vartheta_1 (\theta i t/2)} \frac{\vartheta_1^\prime (0)}{\vartheta_1 (xit/2)}\bigg] \nn\\
=~& -4 \pi \ap l^2 \partial_x \bigg[ \frac{\vartheta_1 (x + \theta_{ab}^i )}{\vartheta_1 (\theta_{ab}^i )} \frac{\eta^3 (2il)}{\vartheta_1 (x)} \bigg].
\end{align}
Now consider the behavior as $l \rightarrow \infty$:
\begin{align}
G_{\theta_{ab}^i} (xit/2) ~\rightarrow~ -\frac{2 \pi \ap l^2}{\sin (\pi \theta )} \partial_x \bigg[ \frac{\sin \pi (x + \theta_{ab}^i )}{ \sin (\pi x)} \bigg].
\end{align}
Thus
\begin{align}
\int_0^{it/2} dz \chi(z) G_{\theta_{ab}^i} (xit/2) ~\rightarrow~& -\frac{i}{2l}   \frac{2 \pi \ap l^2}{\sin (\pi \theta_{ab}^i )} \int_0^1 dx |\sin \pi x|^{-2\ap k^2}   \partial_x \bigg[ \frac{\sin \pi (x + \theta_{ab}^i )}{ \sin (\pi x)} \bigg] \nn\\
~&=~ 2\pi^2 i (\ap)^2 k^2 l  + ...
\end{align}

We can therefore write 
\begin{align}
\A_{\Sigma^i \ov{\Sigma}^i}^X =& -\frac{g^2}{2} \int dl \frac{1}{16\pi^4 (\ap)^2} (2l)^{-3/2} (8\tilde{Z}^\infty)  (2l)^{1/2} 2\sin (\pi\epsilon ) \bigg(\prod_{\kappa=1}^3 \sin (\pi (\theta_{ab}^\kappa - \epsilon))\bigg)\nn\\
&~~~~~~~~~~~\times
\bigg[ 4 \pi^2 i (\ap)^2 k^2 (2l) \bigg] \nn\\
=&-ig^2 K  \frac{4}{16\pi^3 \ap}  (8\tilde{Z}^\infty)   \sin (\pi\epsilon ) \bigg(\prod_{\kappa=1}^3 \sin (\pi (\theta_{ab}^\kappa - \epsilon))\bigg).
\end{align}

\subsection{Total closed string tadpole}

Adding the bosonic and fermionic tadpole contributions, we obtain 
\begin{align}
\A^{tot}_{\Sigma^i \ov{\Sigma}^i} =& -i g^2 K \frac{1}{16\pi^3 \ap} (8\tilde{Z}^\infty)  \bigg[ 4\sin \pi (\theta_{ab}^j - \epsilon) \sin \pi (\theta_{ab}^k - \epsilon) \cos \pi \theta_{ab}^i \nn\\
&\hskip 3.8cm - 2 \sin \pi \theta_{ab}^j \sin \pi \theta_{ab}^k \cos \pi \theta_{ab}^i \nn\\
&\hskip 3.8cm + 4\sin \pi \epsilon \sin \pi (\theta_{ab}^i - \epsilon) \sin \pi (\theta_{ab}^j- \epsilon) \sin \pi (\theta_{ab}^k- \epsilon) \bigg] \nn\\
=& -i g^2 K \frac{1}{16\pi^3 \ap} (8\tilde{Z}^\infty) \bigg[ -1 -\cos^2 \pi \theta_{ab}^i + \cos^2 \pi \theta_{ab}^{j\ne i} + \cos^2 \pi \theta_{ab}^{k\ne j \ne i} \bigg].
\label{total}
\end{align}
Recalling that $8\tilde{Z}^\infty = \prod_{\kappa=3}^3 \frac{\V_a^\kappa \V_b^\kappa}{T_2^\kappa}$ we can match the above to a supergravity calculation, which shall be done exactly in section \ref{SEC:SUGRA}.

\subsection{R-R tadpole cancellation equals two-point tadpole cancellation for supersymmetric amplitudes}

Here we will examine the tadpoles in the supersymmetric case in order to check their cancellation, as we expect from consistency of the theory. We shall follow the approach of \cite{Lust:2003ky}\footnote{A related cancellation occurs in the same models for chiral matter states \cite{Abel:2005qn}.} and consider the explicit case of the $Z_2 \times Z_2$ orientifold. The full tadpole contribution is
\begin{align}
A_{ab} ~\propto~& N_b I_{ab} \cot(\pi \theta^i_{ab}).
\end{align}
The m\"obius strip contribution to the above is
\begin{align}
A_{a, \Omega R g} ~\propto- 4N_a \rho_{\Omega R g} I_{aO6g} \cot(\pi \theta_{a, O6g}^i)
\end{align}
where $g$ is an element of any orbifold group, so $I_{aO6g}^j$ is the intersection number of brane $a$ with the $O6g$ plane in the $j^{\mathrm{th}}$ torus and $I_{aO6g} = I_{aO6g}^1 I_{aO6g}^2 I_{aO6g}^3$. Let us simplify and consider rectangular tori and the $Z_2 \times Z_2$ orbifold. Then there are four group elements: $\{1, \theta, \omega, \theta \omega \}$. Let us take $i=3$. Then 
\begin{align}
A_{ab}^2 ~=~A_{a'b'}^2 ~\propto~& N_b (n_a^1 m_b^1 - n_b^1 m_a^1) (n_a^2 m_b^2 - n_b^2 m_a^2) (\frac{R_1^3}{R_2^3} n_a^3 n_b^3 + \frac{R_2^3}{R_1^3}m_a^3 m_b^3)  \nn\\
A_{ab'}^2 ~=~ A_{a'b}^2 ~\propto~& N_b (-n_a^1 m_b^1 - n_b^1 m_a^1) (-n_a^2 m_b^2 - n_b^2 m_a^2) (\frac{R_1^3}{R_2^3} n_a^3 n_b^3 - \frac{R_2^3}{R_1^3}m_a^3 m_b^3)  \nn\\
&~= N_b (n_a^1 m_b^1 + n_b^1 m_a^1) (n_a^2 m_b^2 + n_b^2 m_a^2) (\frac{R_1^3}{R_2^3} n_a^3 n_b^3 - \frac{R_2^3}{R_1^3}m_a^3 m_b^3)  \nn\\
A_{aa'}^2 ~\propto~& 4 N_a n_a^1 m_a^1 n_a^2 m_a^2 (\frac{R_1^3}{R_2^3} n_a^3 n_b^3 - \frac{R_2^3}{R_1^3}m_a^3 m_b^3)\nn\\
A_{a, \Omega R}^2 ~\propto~& -32  \rho_{\Omega R} m_a^1 m_a^2 n_a^3 \frac{R_1^3}{R_2^3} \nn\\
A_{a, \Omega R \theta}^2 ~\propto~& -32  \rho_{\Omega R \theta} n_a^1 n_a^2 n_a^3 \frac{R_1^3}{R_2^3} \nn\\
A_{a, \Omega R \omega}^2 ~\propto~& 32  \rho_{\Omega R \omega} m_a^1 n_a^2 m_a^3 \frac{R_2^3}{R_1^3} \nn\\
A_{a, \Omega R \theta\omega}^2 ~\propto~& 32  \rho_{\Omega R \theta\omega} n_a^1 m_a^2 m_a^3 \frac{R_2^3}{R_1^3} \nn\\
\end{align}
Thus in total
\begin{align}
A_{UV} &= \sum_{b\ne a} 2 A_{ab} + 2 A_{ab'} + 2A_{aa'} + 2 (A_{a, \Omega R} + A_{a, \Omega R \theta} + A_{a, \Omega R \omega} + A_{a, \Omega R \theta\omega}) \nn\\
&~\propto~~~  4\frac{R_1^3}{R_2^3} \bigg[ m_a^1 m_a^2 n_a^3 \Big( -16 \rho_{\Omega R} + \sum_{b} N_b n_b^1 n_b^2 n_b^3 \Big) \nn\\
&\hskip 4cm +n_a^1 n_a^2 n_a^3 \Big(-16 \rho_{\Omega R \theta} + \sum_{b} N_b m_b^1 m_b^2 n_b^3 \Big) \bigg] \nn\\
&~~~~ +4\frac{R_2^3}{R_1^3} \bigg[ m_a^1 n_a^2 m_a^3 \Big(16  \rho_{\Omega R \omega}-\sum_{b}N_b n_b^1 m_b^2 m_b^3 \Big) \nn\\
&\hskip 4cm + n_a^1 m_a^2 m_a^3 \Big(16  \rho_{\Omega R \theta\omega} -\sum_{b}N_b m_b^1 m_b^2 m_b^3 \Big)\bigg]\nn\\
&=~0
\end{align}
where the vanishing is due to tadpole cancellation, since $\rho_{\Omega R \theta\omega} = \rho_{\Omega R \omega}=\rho_{\Omega R \theta}=-1, \rho_{\Omega R}=1$ and 
\begin{align}
\sum_{b} N_b n_b^1 n_b^2 n_b^3 ~=&~~~~ 16 \nn\\
\sum_{b} N_b m_b^1 m_b^2 n_b^3 ~=& -16 \nn\\
\sum_{b}N_b n_b^1 m_b^2 m_b^3 ~=& -16\nn\\
\sum_{b}N_b m_b^1 m_b^2 m_b^3 ~=& -16.
\end{align}

\subsection{Closed string channel contribution for parallel branes}

It is clear that for the scalars $\Sigma$ in the adjoint representation considered above, the same formula dictates the $\Sigma \ov{\Sigma}$ tadpoles in both tori with parallel and non-parallel branes. However, for the $\Sigma \Sigma$ amplitude, in principle there could have been be other contributions. Clearly there cannot be any from the fermionic pieces, as the correlator is identically zero. There remains the quantum and classical parts of the bosonic correlator $\bra \partial X \partial X \ket$. Firstly the quantum parts are identical for the real and imaginary components, so they both contribute only to the $\Sigma \ov{\Sigma}$ mass. This leaves the classical parts, which comprise the field-theory contribution.

\section{Non-stringy contributions to adjoint scalar masses}
\label{SEC:NONSTRINGY}

Here we shall evaluate the masses in directions where $\theta_{ab}^i =0$, which can be understood as masses from the field theory and not as tadpoles. This will involve the field theory limit of string amplitudes; for more information about this procedure see appendix \ref{APPENDIX:FT} and, for example, \cite{DiVecchia:1996uq}. As we pointed out previously, there cannot be any infra-red poles in the amplitude, and so these must come entirely from the bosonic correlator. Hence we have
\begin{align}
\A_{\Sigma^i \Sigma^i} ~=~& \frac{g^2}{2} \int \frac{dt}{t^2} \frac{1}{16\pi^4 (\ap)^2} \int_0^{it/2} dz \chi(z) \bigg[ 4 \bra \partial X^i (z) \partial X^i (0) \ket_{cl}  \bigg] \nn\\
\A_{\Sigma^i \ov{\Sigma}^i} ~\rightarrow~& \frac{g^2}{2} \int \frac{dt}{t^2} \frac{1}{16\pi^4 (\ap)^2} \int_0^{it/2} dz \chi(z)\bigg[ 4\bra \partial X^i (z) \partial \ov{X}^i (0) \ket \bigg].
\end{align}

We compute (noting the extra sign from the partition function)
\begin{align}
\C{A}_{\Sigma\Sigma}=& - \frac{2g^2}{16\pi^4 (\ap)^2}\int \frac{dt}{t^2} \frac{\vt (\epsilon it/2)^2 \vt ((\theta + \epsilon)it/2) \vt ( (-\epsilon - \theta)it/2)}{\eta^6 (it/2) \vt ((\theta + 2\epsilon)it/2) \vt (( -\theta)it/2)} \nn\\
&\hskip 2.5 cm \times \int_0^{it/2} dz \langle \partial X (z) \partial X(0) \rangle \nn\\
~&\rightarrow~  \frac{2\pi^2 \epsilon^2 g^2}{16\pi^4 (\ap)^2}\int dt \int_0^{it/2} dz \langle \partial X (z) \partial X(0) \rangle_{qu} + \langle \partial X (z) \partial X(0) \rangle_{cl} \nn\\
&=~\frac{2\pi^2 \epsilon^2 g^2}{16\pi^4 (\ap)^2}\int dt \sum_{n_i, m_i} \bigg[  \pi^2 it  \bigg(  n_i^2 \frac{(\ap)^2}{\V_i^2} - \bigg( m_i \frac{T_2^i}{\V_i} + \frac{y_i}{2\pi}\bigg)^2 \bigg) \bigg] \nn\\
&\hskip 2.5 cm \times \exp \bigg[ - \frac{\pi \ap t}{\V_i^2} \bigg| n_i  + i \bigg( m_i \frac{T_2^i}{\ap} + \frac{y_i \V_i}{2\pi \ap}\bigg) \bigg|^2 \bigg] \nn\\
&=~i\frac{2\pi^2 \epsilon^2 g^2}{16\pi^4 (\ap)^2} \sum_{n_i, m_i} \bigg[ \frac{  \bigg(  n_i^2 \frac{(\ap)^2}{\V_i^2} - \bigg( m_i \frac{T_2^i}{\V_i} + \frac{y_i}{2\pi}\bigg)^2}{\bigg(  n_i^2 \frac{(\ap)^2}{\V_i^2} + \bigg( m_i \frac{T_2^i}{\V_i} + \frac{y_i}{2\pi}\bigg)^2 \bigg)^2} \bigg].
\end{align}
If we now examine the zero mode where $n_i = m_i =0$ (which dominates the amplitude) we have
\begin{align}
\C{A}_{\Sigma\Sigma} \rightarrow & -i \frac{2\pi^2 \epsilon^2 g^2}{16\pi^4 (\ap)^2}\int dt \; t \frac{y^2}{4} \exp [ - t\frac{y^2}{4\pi\ap} ]\nn\\
&= -i \frac{2 \epsilon^2 g^2}{64\pi^2 (\ap)^2}\int dt \; t y^2 \exp [ - t\frac{y^2}{4\pi\ap} ]\nn\\
&=  -i \frac{2 \epsilon^2 g^2}{64\pi^2 (\ap)^2}\int dt \; t 4\pi^2 (\ap)^2 M_0^2 \exp [ - t \pi \ap M_0^2]\nn\\
&= -i \frac{2 \epsilon^2 g^2}{16\pi^2 (\ap)^2} \frac{1}{M_0^2} 
\end{align}
which is the correct result according to the effective potential calculation. Finally we compute
\begin{align}
\C{A}_{\Sigma\ov{\Sigma}}~\rightarrow& ~ \frac{2\pi^2 \epsilon^2 g^2}{16\pi^4 (\ap)^2}\int dt \int_0^{it/2} dz \langle \partial X (z) \partial \ov{X}(0) \rangle_{qu} + \langle \partial X (z) \partial \ov{X}(0) \rangle_{cl} \nn\\
&=~\frac{2\pi^2 \epsilon^2 g^2}{16\pi^4 (\ap)^2}\int dt \sum_{n_i, m_i} \bigg[ -i\pi \ap +  \pi^2 it  \bigg(  n_i^2 \frac{(\ap)^2}{\V_i^2} + \bigg( m_i \frac{T_2^i}{\V_i} + \frac{y_i}{2\pi}\bigg)^2 \bigg) \bigg] \nn\\
&~~~~~~~~~~~~\times \exp \bigg[ - \frac{\pi \ap t}{\V_i^2} \bigg| n_i  + i \bigg( m_i \frac{T_2^i}{\ap} + \frac{y_i \V_i}{2\pi \ap}\bigg) \bigg|^2 \bigg] \nn\\
&=~0.
\end{align}
Examining only the zero mode, we have
\begin{align}
\C{A}_{\Sigma\ov{\Sigma}} ~\supset~& \frac{2\pi^2 \epsilon^2 g^2}{16\pi^4 (\ap)^2}\int dt \bigg[ -i\pi \ap +  \pi^2 it   \frac{y_i^2}{4\pi^2} \bigg] \exp [ - t\frac{y^2}{4\pi\ap} ] \nn\\
~&=~i \frac{ \epsilon^2 g^2}{32\pi^2 (\ap)^2}\int dt \bigg[ -4\pi \ap +  t y_i^2 \bigg] \exp [ - t\frac{y^2}{4\pi\ap} ].
\end{align}
This exactly matches the effective potential calculation.

\section{Masses beyond the leading order}
\label{SEC:MASSESBEYONDLEADING}

As observed above in section \ref{SEC:CW}, the Coleman-Weinberg IR masses are traceless at the leading order. This can be seen as a result of the following:
\begin{align}
\mathrm{tr} (m^2) ~=~& i \sum_i \A_{\Sigma^i \ov{\Sigma}^i} \nn\\ 
~=~& i\frac{g^2}{2} \sum_i  \int \frac{dt}{t^2} \frac{1}{16\pi^4 (\ap)^2} \int_0^{it/2} dz \chi(z)\bigg[ 4\bra \partial X^i (z) \partial \ov{X}^i (0) \ket \bigg] \nn\\
~=~&  i2 g^2 \sum_i  \int \frac{dt}{t^2} \frac{1}{16\pi^4 (\ap)^2} \bigg(\sum_\nu \frac{\delta_\nu}{2} Z_\nu (it/2) \bigg)  \times \nonumber \\
&~~~~~ \bigg[ -i\pi \ap +  \pi^2 it  \bigg(  n_i^2 \frac{(\ap)^2}{\V_i^2} + \bigg( m_i \frac{T_2^i}{\V_i} + \frac{y_i}{2\pi}\bigg)^2 \bigg) \bigg] \prod_\kappa Z_{cl}^\kappa \nn\\
~=~& 2 \pi \ap g^2 \sum_i  \int \frac{dt}{t^2} \frac{1}{16\pi^4 (\ap)^2} \bigg(\sum_\nu \frac{\delta_\nu}{2} Z_\nu (it/2) \bigg) \bigg[ Z_{cl}^i + t \frac{d}{dt} Z_{cl}^i \bigg] \prod_{\kappa \ne i} Z_{cl}^\kappa \nn\\
~=~&  \frac{g^2}{8\pi^3 \ap}\int dt \bigg\{ \frac{1}{t^{1+n_{\parallel}}} \sum_\nu \frac{\delta_\nu}{2} Z_\nu (it/2) \bigg\} \frac{d}{dt} \bigg(  \prod_{\kappa=1}^{n_\parallel} t Z_{cl}^\kappa \bigg) ,
\end{align}
where $n_\parallel$ is the number of parallel directions, equal to $2$ for the $N\approx 4$ case and $0$ or $1$ for the $N\approx 2$ cases. Now $Z_{cl}^\kappa \rightarrow 0$ as $t\rightarrow \infty, 0$, in both cases exponentially, so we see that in the limit that the factor in curly brackets is a constant, (i.e. the leading order term!) then the integral vanishes. However, the subleading order in $\epsilon$ term will not in general be a constant, and so we expect a non-zero contribution from the integral. For example, in the $N\approx 4$ case we have
\begin{align}
\frac{1}{t^{1+n_\parallel}} \sum_\nu \frac{\delta_\nu}{2} Z_\nu (it/2) ~=~& \frac{i}{t^3} \frac{\vt (\epsilon it/2)^4}{\vt(\epsilon it) \eta^9 (it/2)} I_{ab} \nn\\
~=~& \frac{i}{t^3} I_{ab}\bigg[ (2\pi)^3 \frac{( \vt^\prime (0) \epsilon it/2 + \frac{1}{6} \vt^{\prime\prime \prime} (0) (\epsilon it/2)^3 + ...)^4}{( \vt^\prime (0) \epsilon it + \frac{1}{6} \vt^{\prime\prime \prime} (0)  (\epsilon it)^3 + ...)\vt^\prime (0)^3 } \bigg] \nn\\
~=~& |\epsilon|^3 |I_{ab}|\frac{\pi^3}{2} \bigg[ 1 - \frac{\epsilon^4 t^4}{8} \bigg(\frac{2}{5!} \vt^\prime(0) \vt^{\prime\prime\prime\prime \prime} (0) - \left(\frac{\vt^{\prime\prime\prime}(0)}{3!}\right)^2\bigg) + ... \bigg] 
\end{align} 
So we would expect there to be a contribution to the trace at order $\epsilon^7$. These do involve string oscillators in the loop and so could not be seen from the field theory, but unfortunately appear at such a subleading order that it is doubtful that they may be of phenomenological use.

\section{Supergravity derivation}
\label{SEC:SUGRA}

In this section we shall compute the UV divergences appearing in the two-point function for adjoint scalars via effective supergravity, demonstrating that they are in fact due to the presence of NS-NS tadpoles. Note that in this section we shall take our metric conventions to be $\eta^{\mu\nu} = \mathrm{diag}(-1, 1, 1, 1,...)$.

What we shall calculate is illustrated as follows. 
Consider a toy Lagrangian:
\begin{align}
\C{L} ~\supset~& - \frac{1}{2}( \partial_\mu \psi \partial^\mu \psi + \partial_\mu \phi \partial^\mu \phi) - a\phi \partial_\mu \psi \partial^\mu \psi - b \phi \nn\\
~&\rightarrow~ a\phi (k_3) k_1 \cdot k_2 \psi (k_1) \psi (k_2)\delta(k_1 + k_2 + k_3) - b \phi (k_3)\delta (k_3)
\end{align} 
Since this contains a tadpole, $\phi=0$ is no longer a solution of the equations of motion, so we are working in a false vacuum \cite{Tadpoles}. However, if we persist with the above theory then with a propagator of $-i/k_3^2$ for $\phi$ we generate the amplitude for $\bra \psi \psi \ket$ of
\begin{align}
\C{A} ~=~& 2(iak_1 \cdot k_2) (-ib) \bigg[\frac{-i}{k_3^2}  \bigg]_{k_3=0} = 2(-iak_1^2) (-ib) \bigg[\frac{-i}{k_3^2}  \bigg]_{k_3=0} \nn\\
~=~&2i ab k_1^2  \bigg[\frac{1}{k_3^2} \bigg]_{k_3=0}.
\end{align} 
i.e. the term in square brackets is divergent. To match the factors with the closed string calculation, imagine regulating the above by adding a mass term for the field $\phi$; then we would have
\begin{align}
\C{A} ~\rightarrow~&  2i ab k_1^2 \bigg[\frac{1}{k_3^2 + M^2} \bigg]_{k_3=0} \nn\\
~\rightarrow~&  \frac{2i ab k_1^2}{M^2} \nn\\
&=~ 2i ab k_1^2 \int_0^\infty dl^\prime e^{-M^2 l^\prime},
\end{align} 
where we have written the last line in a suggestive form. To match this to the string computation, recalling that the partition function in the closed string channel contains $\exp [ - \pi l (L_0 + \tilde{L}_0)] $ and so we can write $l^\prime = \pi \ap l$ and
\begin{align}
\C{A} ~\rightarrow~& 2 i ab k_1^2 \int_0^\infty dl \pi \ap e^{-\pi \ap M^2 l} \nn\\
~\equiv~& -i A_{UV} \tilde{K},
\end{align}
where $\tilde{K} = -K$, the latter being defined in (\ref{DefineK}). The relative minus sign is to account for the different metrics we are using, so that the coefficients will be the same; in this section we shall determine the coefficient $\C{A}_{UV}$ from a supergravity calculation.

Let us first derive the relationship between coordinates and adjoints. We start with the action for the dilaton, two-form and graviton:
\begin{align}
S_{NS} ~=~& \frac{1}{2\kappa_{10}^2} \int d^{10} x \sqrt{-G} e^{-2\Phi} ( R + 4 \partial_\mu \Phi \partial^\mu \Phi - \frac{1}{2} |H_3|^2)
\end{align}
where $2\kappa_{10}^2 = (\ap)^4 (2\pi)^7$. Then the DBI action for D$p$-branes in the string frame:
\begin{align}
S_{DBI} = -2\pi (4\pi^2\ap)^{-(p+1)/2} \int d^{p+1} x e^{-\Phi} \sqrt{-\mathrm{det} (g + B + 2\pi \ap F)}.
\end{align}
To perform the supergravity calculation, however, it is convenient to transform to the Einstein frame to separate the graviton and dilaton actions, by writing $G = e^{\Phi/2} G_E$ so that the action becomes 
\begin{align}
S_{NS} =&~ \frac{1}{2\kappa_{10}^2} \int d^{10} x \sqrt{-G_E}( R_E - \frac{1}{2} \partial_\mu \Phi \partial^\mu \Phi - \frac{1}{2} e^{-\Phi}|H_3|^2) \nn\\
S_{DBI}=& -2\pi (4\pi^2\ap)^{-(p+1)/2} \int d^{p+1} x e^{(p-3)\Phi/4} \sqrt{-\mathrm{det} (g_E + e^{-\Phi/2 }B + \ell e^{-\Phi/2}F)}.
\end{align}

Now put the components of the gauge field tangential to the brane as $x^\alpha$, and the transverse fluctuations as  $\zeta^n$ where $n$ labels an index normal to the brane. The pullback of the metric to the brane world-volume is 
\begin{align}
\phi^* (g)_{\mu \nu} ~=~& g_{\mu \nu} + \ell g_{\mu n} \nabla_\nu \zeta^n + \ell g_{\nu n} \nabla_\mu \zeta^n  + \ell^2 g_{n m} \nabla_\mu \zeta^n \nabla_\nu \zeta^m  + ...
\end{align}
where the connection is in the normal bundle of the world-volume and $\mu$ is an index tangential to the brane, and $\ell \equiv 2\pi \ap$. Since we are dealing with tori, the connection is flat. We are only interested in the derivatives in the 4d directions, the adjoints only exist in the compact ones; write $\mu, \nu$ for 4d indices, $\alpha, \beta$ for tangential ones and $m,n$ for perpendicular, and since we are comparing to a string computation, we should not rescale the fluctuations, then
\begin{align}
\phi^* (g)_{\mu \nu} ~=~& e^{\Phi/2}\Big[\eta_{E\;\mu\nu} + h_{\mu\nu}+  \ell^2 g_{E\; n m} \partial_\mu \zeta^n \partial_\nu \zeta^m  + \ell^2 h_{nm}\partial_\mu \zeta^n \partial_\nu \zeta^m + ...  \Big] \nn\\
\phi^* (g)_{\mu \beta} ~=~& e^{\Phi/2} \ell h_{n\beta } \partial_\mu \zeta^n +... \nn\\
\phi^* (g)_{\alpha \nu} ~=~& e^{\Phi/2} \ell h_{\alpha m} \partial_\nu \zeta^m +... \nn\\
\phi^* (g)_{\alpha \beta} ~=~&e^{\Phi/2}\Big[ g_{E\; \alpha \beta} + h_{\alpha \beta} + ... \Big]
\end{align}

Then using
\begin{align}
\sqrt{-\det (g + X)} = \sqrt{-\det g} \bigg[& 1 + \frac{1}{2} \tr(g^{-1} X) + \frac{1}{8} \bigg(\tr(g^{-1} X)\bigg)^2 - \frac{1}{4} \tr\bigg( (g^{-1} X)^2\bigg) \nn\\
&- \frac{1}{8} \tr(g^{-1} X) \tr\bigg( (g^{-1} X)^2\bigg) + \frac{1}{6} \tr\bigg((g^{-1} X)^3\bigg) + ... \bigg],
\end{align}
the couplings of the normal directions are given by 
\begin{align}
S_{\zeta}=& -2\pi \ell^2 (4\pi^2\ap)^{-(p+1)/2} e^{(p-3)\bra \Phi\ket /4} \nn\\
&\times \int d^{p+1} x \sqrt{-\det g_E} \bigg[ \frac{1}{2} g_{n m} \partial_\mu \zeta^n \partial^\mu \zeta^m %\nn\\
%&
+ \frac{p-3}{8}\Phi g_{n m} \partial_\mu \zeta^n \partial^\mu \zeta^m  + \frac{1}{2} h_{n m} \partial_\mu \zeta^n \partial^\mu \zeta^m  \nn\\
&\hskip 1.5cm - \frac{1}{2} \eta^{\rho \mu} g_{E\; n m} \partial_\mu \zeta^n \partial_\nu \zeta^m \eta^{\nu \lambda} h_{\lambda \rho} + \frac{1}{4}(g^{\alpha \beta} h_{\beta \alpha}) g_{n m} \partial_\mu \zeta^n \partial^\mu \zeta^m\bigg].
\end{align}
We can expand the rest of the DBI action to 
\begin{align}
S_{DBI}~=& -2\pi (4\pi^2\ap)^{-(p+1)/2}  \nn\\
&~~~~\times \int d^{p+1} x \sqrt{-\det g_E} \Big[  e^{(p-3)\bra\Phi \ket/4} \Big(1 + \frac{p-3}{4} \Phi + \eta^{\mu\nu} h_{\mu\nu} + g^{\alpha \beta} h_{\alpha\beta} \Big)\nn\\
& \hskip 2.5cm + \frac{\ell^2}{4} e^{\frac{p-7}{4}\bra \Phi\ket} \Big(1 + \frac{p-7}{4}\Phi + \frac{1}{2} \eta^{\mu\nu} h_{\mu\nu} + \frac{1}{2} g^{\alpha \beta} h_{\alpha \beta}\Big) F_{\mu\nu} F^{\mu\nu} \nn\\
&\hskip 2.5cm + \frac{\ell^2}{2} e^{\frac{p-7}{4}\bra \Phi\ket} h_{\mu \nu} F^{\nu \rho} F_{\rho \lambda} \eta^{\lambda \mu}  \Big].
\end{align}

For a flat background, the kinetic terms are given by
\begin{align}
S_{NS} = - \frac{1}{8\kappa_{10}^2} \int d^{10} x \bigg( \partial_\mu h_{\nu\lambda} \partial^\mu h^{\nu\lambda} - \frac{1}{2}\partial_\mu h^\nu_\nu \partial^\mu h^\lambda_\lambda + 2 \partial_\mu \Phi \partial^\mu \Phi\bigg) .
\end{align}
The corresponding propagators are given by
\begin{align}
\bra h_{\mu\nu} h_{\sigma \rho}\ket =& - \frac{2i\kappa_{10}^2}{k^2} \bigg( \eta_{\mu \sigma}\eta_{\nu \rho} +  \eta_{\mu \rho}\eta_{\nu \sigma} - \frac{2}{d-2} \eta_{\mu\nu} \eta_{\sigma \rho}\bigg) \nn\\
\bra \Phi \Phi \ket =& - \frac{2i\kappa_{10}^2}{k^2}.
\end{align}
 It is then straightforward to show that, if we had a true 4d graviton, its contribution to gauge boson and adjoint tadpoles is zero: restrict the indices to only the non-compact dimensions, then the operator in the effective potential is proportional to 
\begin{align}
&\bigg[\frac{1}{4}  (F_{\mu'\nu'} F^{\mu'\nu'} + 2 F_{\mu'\alpha'} F^{\mu'\alpha'}) \eta^{\mu \nu} + (F^{\nu \rho} F_{\rho\lambda} + 2 F^{\nu \alpha} F_{\alpha \lambda} ) \eta^{\lambda\mu} \bigg] \nn\\
&\hskip 6.5cm\times \bigg( \eta_{\mu \sigma}\eta_{\nu \rho} +  \eta_{\mu \rho}\eta_{\nu \sigma} - \eta_{\mu\nu} \eta_{\sigma \rho}\bigg) \eta^{\sigma\rho}\nn\\
&~~~~~~~~~~~~=~ \bigg[\frac{1}{4}  (F_{\mu'\nu'} F^{\mu'\nu'} + 2 F_{\mu'\alpha'} F^{\mu'\alpha'}) \eta^{\mu \nu} + (F^{\nu \rho} F_{\rho\lambda} + 2 F^{\nu \alpha} F_{\alpha \lambda} ) \eta^{\lambda\mu} \bigg] (-2 \eta_{\mu\nu})\nn\\
&~~~~~~~~~~~~=~ \bigg[ (F_{\mu'\nu'} F^{\mu'\nu'} + 2 F_{\mu'\alpha'} F^{\mu'\alpha'})  + (F^{\nu \rho} F_{\rho\nu} + 2 F^{\nu \alpha} F_{\alpha \nu} )  \bigg]\nn\\
&~~~~~~~~~~~~=~ 0.
\end{align}
Clearly the 4d graviton does not couple to the brane tadpole. Since this is the only field that could mediate masses after moduli stabilization, we can see that all adjoint scalar tadpoles generated in this way vanish once the moduli are made massive. However, we can still check the results in the non-stabilised case, where the 4d and compact components mix! There, we must retain $d=10$ in the propagator, which encapsulates the mixing of the graviton with the moduli.

The metric on a 2-torus is given by $\frac{T_2}{U_2} | dx + U dy|^2$, where $T = T_1 + i T_2 = i R_1 R_2 \sin \alpha$, $U = U_1 + iU_2 = \frac{R_2}{R_1} e^{i\alpha}$. A brane wraps a cycle defined by $(x, y) = (2\pi n\lambda, 2\pi m \lambda)$ where $\lambda \in [0,1]$. Then we need to study the normal direction; a vector in the tangent bundle is $(n,m)$ and we require $g_{\alpha n} = 0$. Let us for simplicity choose rectangular tori, $\alpha = \pi/2$. Then $g_{11} = R_1^2$, $g_{22} = R_2^2$. Let us also normalise the normal direction so that the coordinate $\zeta \in [0,1]$; the normal direction is $2\pi  \frac{\zeta}{n^2 R_1^2 + m^2 R_2^2} (-mR_2^2, nR_1^2) = 4\pi^2 \frac{\zeta}{L^2} (-mR_2^2,~ nR_1^2)$, or $2\pi \frac{\zeta}{\C{V}^2} (-mR_2^2, nR_1^2)$ where $L = 2\pi \C{V}$. Note that the normal distance between the branes is $4\pi^2 R_1 R_2/L$.

Thus the graviton pieces tangential and normal to a brane are
\begin{align}
h_{\lambda\lambda} ~=~& 4\pi^2  \bigg[ n^2 h_{11} + nm (h_{12} + h_{21}) + m^2 h_{22} \bigg]\nn\\
h_{nn} ~=~&\frac{4\pi^2}{\V^4} \bigg[ m^2 R_2^4 h_{11} -nm R_1^2 R_2^2 (h_{12} + h_{21}) + n^2 R_1^4 h_{22} \bigg]  
\end{align}
The propagators become
\begin{align}
\bra h_{11} h_{11}\ket ~=~& - \frac{2i\kappa_{4}^2}{k^2} \frac{7}{4} R_1^4 \nn\\
\bra h_{22} h_{22}\ket ~=~& - \frac{2i\kappa_{4}^2}{k^2} \frac{7}{4} R_2^4 \nn\\
\bra h_{12} h_{12}\ket ~=~& - \frac{2i\kappa_{4}^2}{k^2}  R_1^2 R_2^2  = \bra h_{12} h_{21}\ket\nn\\
\bra h_{11} h_{12}\ket ~=~& 0 ~=~ \bra h_{22} h_{12}\ket .
\end{align}

We can write the couplings of the adjoints to the moduli:
\begin{align}
F_{\mu \nu} F^{\mu \nu } ~\rightarrow~& F_{\mu \nu} F^{\mu \nu } + 2F_{\mu \lambda} F^{\mu \lambda}\nn\\
~=~&  -  \frac{h_{\lambda \lambda}}{(4\pi^2) (n^2 R_1^2 + m^2 R_2^2)}  \bigg(4\pi^2 (n^2 R_1^2 + m^2 R_2^2)\eta^{\mu\nu} \partial_\mu A^\lambda \partial_\nu A^\lambda\bigg)
\end{align}
Now in the string computation we normalise the adjoints so that their kinetic terms are that of the gauge coupling
\begin{align}
S ~\supset~&  -2\pi (4\pi^2\ap)^{-(p+1)/2}  \nn\\
&~~~~~~~~~~\times \int d^{p+1} x \sqrt{-\det g_E} \bigg[ \frac{\ell^2}{4} e^{\frac{p-7}{4}\bra \Phi\ket} F_{\mu\nu} F^{\mu\nu} + \frac{\ell^2}{2} e^{\frac{p-7}{4}\bra \Phi\ket} h_{\mu \nu} F^{\nu \rho} F_{\rho \lambda} \eta^{\lambda \mu} \bigg]\nn\\
~=~& 2\pi (4\pi^2\ap)^{-(p+1)/2} \ell^2 e^{\frac{p-7}{4}\bra \Phi\ket} L_a \int d^4 x - \frac{1}{4} F_{\mu\nu} F^{\mu\nu} +...
\end{align}
So the gauge coupling is
\begin{align}
\frac{1}{g^2} ~=~  2\pi (4\pi^2\ap)^{-(p+1)/2} \ell^2 e^{\frac{p-7}{4}\bra \Phi\ket} L_a.
\end{align}
Meanwhile the coupling for the transverse adjoints is
\begin{align}
S~\supset~ &2\pi (4\pi^2\ap)^{-7/2} \ell^2 e^{\frac{p-7}{4}\bra \Phi\ket}  \int d^7 x \sqrt{-\det g}  - \frac{1}{2} g_{n m} \partial_\mu \zeta^n \partial^\mu \zeta^m 
\end{align}
Then we define for a torus $\kappa$ the complex adjoint to be composed of the gauge field $A^\kappa$ and the normal coordinate $\zeta^\kappa$ via
\begin{align}
\Sigma^\kappa ~\equiv~& \sqrt{2}\pi \bigg[ \sqrt{n^2 R_1^2 + m^2 R_2^2} A^\kappa + i\frac{R_1 R_2}{\sqrt{n^2 R_1^2 + m^2 R_2^2}} \zeta^\kappa\bigg]
\end{align}
which has kinetic term $-\frac{1}{g^2} |\partial_\mu \Sigma|^2$. Then the couplings of the gauge part of the adjoints to the closed strings are
\begin{align}
S ~\supset~& -\frac{1}{4g^2} \int d^4 x\bigg[ (1 + \frac{p-7}{4}\Phi + \frac{1}{2} \eta^{\mu\nu} h_{\mu\nu} + \frac{1}{2} g^{\alpha \beta} h_{\alpha \beta}) F_{\mu\nu} F^{\mu\nu} + 2 h_{\mu \nu} F^{\nu \rho} F_{\rho \lambda} \eta^{\lambda \mu} \bigg] \nn\\
~\supset~&  -\frac{1}{2g^2} \int d^4 x\bigg[ \sum_{\kappa=1}^3 (\frac{1}{2}\eta^{\mu\nu}\eta^{\rho\omega} - \eta^{\mu\omega}\eta^{\nu\rho})h_{\mu\nu} \partial_\rho A^\kappa \partial_\omega A^\kappa  + (\partial_\mu A_r^\kappa)^2\bigg(\frac{p-7}{4}\Phi  \nn\\
&\hskip 3.5cm -\frac{1}{2}\frac{h_{\lambda \lambda}^\kappa}{(4\pi^2) (n^2 (R_1^\kappa)^2
+ m^2 (R_2^\kappa)^2)} \nn\\
& \hskip 3.5cm + \sum_{j \ne \kappa}\frac{1}{2} \frac{h_{\lambda \lambda}^j}{(4\pi^2)(n^2 (R_1^j)^2 + m^2 (R_2^j)^2) } \bigg)  
\end{align}
where we have now indexed the internal dimensions.

The couplings of the normal directions are  given by
\begin{align}
S_{\zeta}=& -\frac{1}{2g^2} \int d^{4} x \bigg[ g_{n m} \partial_\mu \zeta^n \partial^\mu \zeta^m + \frac{p-3}{4}\Phi g_{n m} \partial_\mu \zeta^n \partial^\mu \zeta^m  +  h_{n m} \partial_\mu \zeta^n \partial^\mu \zeta^m  \nn\\
&\hskip 2cm -  \eta^{\rho \mu} g_{E\; n m} \partial_\mu \zeta^n \partial_\nu \zeta^m \eta^{\nu \lambda} h_{\lambda \rho} + \frac{1}{2}(g^{\alpha \beta} h_{\beta \alpha}) g_{n m} \partial_\mu \zeta^n \partial^\mu \zeta^m\bigg].
\end{align}
This becomes
\begin{align}
S_{\zeta}=& -\frac{1}{2g^2} \int d^{4} x \sum_{\kappa=1}^3 \bigg[ \partial_\mu \zeta^\kappa \partial^\mu \zeta^\kappa + \frac{p-3}{4}\Phi \partial_\mu \zeta^\kappa \partial^\mu \zeta^\kappa + h_{\kappa \kappa} \frac{\V_\kappa^2}{4\pi^2 R_1^\kappa R_2^\kappa}\nn\\
&\hskip 3.5cm -  \partial^\mu \zeta^\kappa \partial^\nu \zeta^\kappa h_{\nu\mu} + \sum_{\lambda=1}^3 \frac{1}{2} \frac{h_{\lambda \lambda}}{4\pi^2 V_\lambda^2} \partial_\mu \zeta^\kappa \partial^\mu \zeta^\kappa \bigg]
\end{align}

Now let us amplitude involving two different branes. These couple to the tadpole
\begin{align}
S ~\supset&   - \frac{1}{g_b^2} e^{\bra\Phi\ket} \int d^4 x \left( 1+ \frac{p-3}{4} \Phi + \frac{1}{2} \eta^{\mu \nu} h_{\mu\nu} + \sum_{k=1}^3 \frac{1}{2}\frac{h_{kk}}{4\pi^2 (\V_b^k)^2}\right).
\end{align}
The contribution of the ``4d graviton'' (note that this is mixed with the moduli) to the amplitude is thus, for the non-compact part of the tadpole
\begin{align}
\C{A} ~\supset~&2\times \frac{e^\Phi}{\ell^2}\frac{-2i\kappa_4^2 }{2g_a^2 g_b^2}\frac{-ik_\rho k_\omega}{k^2} (\frac{1}{2}\eta^{\mu\nu}\eta^{\rho\omega} - \eta^{\mu\omega}\eta^{\nu\rho}) (\eta_{\mu\sigma} \eta_{\nu\rho} + \eta_{\mu \rho} \eta_{\nu \sigma} - \frac{1}{4} \eta_{\mu\nu} \eta_{\sigma \rho})\frac{(-i)}{2} \eta^{\sigma \rho} \nn\\
&=-2\times\frac{e^\Phi}{\ell^2}\frac{-2i\kappa_4^2 }{4g_a^2 g_b^2}\frac{k_\rho k_\omega}{k^2} (\frac{1}{2}\eta^{\mu\nu}\eta^{\rho\omega} - \eta^{\mu\omega}\eta^{\nu\rho}) \eta_{\mu\nu} \nn\\
&=2\times\frac{e^\Phi}{\ell^2}\frac{-2i\kappa_4^2 }{4g_a^2 g_b^2}\tilde{K}~ = ~-i e^\Phi\frac{2\kappa_4^2}{32g_a^2 g_b^2\ell^2} \times 8 \tilde{K}
\end{align}
with also a contribution from the compact part of the tadpole
\begin{align}
\C{A} ~\supset~&2\times\frac{e^\Phi}{\ell^2}\frac{-2i\kappa_4^2 }{2g_a^2 g_b^2}\frac{-ik_\rho k_\omega}{k^2} (\frac{1}{2}\eta^{\mu\nu}\eta^{\rho\omega} - \eta^{\mu\omega}\eta^{\nu\rho}) ( - \frac{1}{4} \eta_{\mu\nu} g_{kk})\frac{(-i)}{2} g^{kk} \nn\\
~&=~2\times\frac{e^\Phi}{\ell^2}\frac{-2i\kappa_4^2 }{4g_a^2 g_b^2} \times \left( -\frac{3}{4}\right) \tilde{K} ~=~  -i e^\Phi\frac{2\kappa_4^2}{32g_a^2 g_b^2\ell^2} \times (-6)\tilde{K}.
\end{align}
The dilaton  gives
\begin{align}
\C{A} ~\supset~& 2\times\bigg(\frac{-2i\kappa_4^2 }{k^2}\bigg) \bigg( -i \frac{p-3}{4 \ell^2 g_b^2} e^{\Phi }\bigg) \bigg(-ik^2 \frac{p-7}{8g_a^2}\bigg) \nn\\
&= - 2\times ie^\Phi \frac{2\kappa_4^2}{32 \ell^2  g_a^2 g_b^2} \times (-3)\tilde{K}.
\end{align}

The compact part of the ``graviton'' within the same torus gives
\begin{align}
&\bra h_{\lambda_a \lambda_a}^\kappa h_{\lambda_b \lambda_b}^\kappa\ket =\nn\\
&~~~(4\pi^2)^2\Big\bra \Big(n^2_a h_{11} + n_am_a (h_{12} + h_{21}) + m^2_a h_{22} \Big)\Big(n^2_b h_{11} + n_bm_b (h_{12} + h_{21}) + m^2_b h_{22} \Big)\Big\ket  \nn\\
&~~~~~~~~~~~~~~~~= - \frac{2i\kappa_{4}^2}{k^2}  (4\pi^2)^2\bigg[ 2(\V_a^\kappa)^2 (\V_b^\kappa)^2 \cos^2 \pi\theta_{ab}^\kappa - \frac{1}{4} (\V_a^\kappa)^2 (\V_b^\kappa)^2 \bigg],
\end{align}
while between different tori we have
\begin{align}
&\bra h_{\lambda_a \lambda_a}^k h_{\lambda_b \lambda_b}^j \ket ~=~(4\pi^2)^2\Big\bra \Big(( n^k_a)^2 h_{11}^k + n_a^km_a^k (h_{12}^k + h_{21}^k) + (m^k_a)^2 h_{22}^k \Big)
\nn\\
&\hskip 4.5cm \times
\Big((n_b^j)^2 h_{11}^j + n_b^jm_b^j (h_{12}^j + h_{21}^j) + (m^j)_b^2 h_{22}^j \Big)\Big\ket  \nn\\
&~~~~~~~~~~~~~~~~~=  \frac{2i\kappa_{4}^2}{k^2}  (4\pi^2)^2 \frac{1}{4}(\V_a^k)^2 (\V_b^j)^2 
\end{align}
and between compact and non-compact directions we have
\begin{align}
\bra h_{\lambda_a \lambda_a}^k h_{\mu \nu} \ket &~=~ (4\pi^2)\Big\bra \Big(n^2_a h_{11} + n_am_a (h_{12} + h_{21}) + m^2_a h_{22} \Big) h_{\mu\nu}\Big\ket  \nn\\
&~=~   \frac{2i\kappa_{4}^2}{k^2}  (4\pi^2) \frac{1}{4} (\V_a^j)^2  \eta_{\mu\nu}.
\end{align}

Thus the total divergence is given by $\C{A} = -i A_{UV} \tilde{K}$ where
\begin{align}
A_{UV} ~=~& \frac{e^\Phi}{16g_a^2 g_b^2 \ell^2} 2\kappa_{4}^2 \bigg[  -8\cos^2 \pi\theta_{ab}^k + 8\cos^2 \pi \theta_{ab}^{j\ne k} + 8\cos^2 \pi \theta_{ab}^{i\ne k \ne j} - 7 - 3 - 6 + 8 \bigg] \nn\\
~=~& \frac{e^\Phi}{2g_a^2 g_b^2\ell^2} 2\kappa_{4}^2 \bigg[  -1 - \cos^2 \pi\theta_{ab}^k + \cos^2 \pi \theta_{ab}^{j\ne k} + \cos^2 \pi \theta_{ab}^{i\ne k \ne j}   \bigg].
\end{align}
(Note however that the physical divergence is given by $g_a^2$ multiplying the above). Let us rearrange this using 
\begin{align}
2\kappa_4^2 ~=~& \frac{(2\pi)^7 (\ap)^4}{(2\pi)^6 T_2^1 T_2^2 T_2^3} \nn\\
\frac{1}{g_b^2}~=~& 2\pi (4\pi^2\ap)^{-(p+1)/2} \ell^2 e^{\frac{p-7}{4}\bra \Phi\ket} (2\pi)^3 \V_b
\end{align}
to obtain
\begin{align}
 A_{UV}~=~& \frac{1}{2\ap} \frac{\V_a \V_b}{(2\pi)^3 T_2^1 T_2^2 T_2^3} e^{ \Phi/2} \bigg[  -1 - \cos^2 \pi\theta_{ab}^k + \cos^2 \pi \theta_{ab}^{j\ne k} + \cos^2 \pi \theta_{ab}^{i\ne k \ne j}   \bigg].
\end{align}
This is the main result of this section, which is in agreement with (\ref{total}).

Note that for completeness we could also compute the result for the normal components. However, we already know from the CFT computation that these will give the same result; while it may be useful to do this straightforward calculation as a check, we shall leave this as an exercise for the reader.

Finally, it is readily shown, using the techniques above, that tapoles are generated for gauge bosons too, given by:
\begin{align}
A_{A_a} ~=~& \frac{1}{2\ap} \frac{\V_a \V_b}{(2\pi)^3 T_2^1 T_2^2 T_2^3} e^{ \Phi/2} \bigg[  -3 + \cos^2 \pi\theta_{ab}^k + \cos^2 \pi \theta_{ab}^{j\ne k} + \cos^2 \pi \theta_{ab}^{i\ne k \ne j}   \bigg].
\end{align}
This is perhaps the clearest indication that these contributions cannot survive in a theory with stabilised moduli.

\section{Conclusions}
\label{SEC:CONCLUSIONS}

The reductions of higher-dimensional gauge fields to four dimensions lead to  massless scalars in adjoint representations  (Wilson lines). These states may, or may not, survive  accompanying projections applied  to  get down from $N=4$ 4d supersymmetry to $N=1$. If they do, then they are expected to acquire masses when supersymmetry is fully broken. 

There are very few classes of string compactifications where such effects can be computed fully and explicitly. We have considered here the case where supersymmetry breaking is obtained when brane intersection angles are deformed away from their special values, corresponding to a supersymmetric configuration, by a small angular shift $2\epsilon$. This leads to supersymmetry breaking via a D-term vacuum expectation value $\bra D\ket $, associated to a magnetised abelian gauge group factor in the T -dual picture. All charged scalar fields localised at the intersections obtain supersymmetry breaking mass shifts, and play the role of mediator messengers.
 
We have written down the one-loop propagator of the  open string states  in adjoint representations and extracted the leading terms at vanishing external momentum.  The result is understood as the sum of two parts. 

The first part comes from the ultraviolet limit in the open string channel, equivalent to the infrared  limit of the exchange of massless closed string states.   It is understood in the tree-level  effective supergravity,  and is shown to correspond to reducible diagrams. It represents the interaction with global tadpoles through emission of the corresponding massless moduli. Such tadpoles should be cancelled in a stable background with the corresponding moduli fixed. Thus, we expect these contributions to be modified in the true vacuum (and be probably vanishing).

The second part describes the effects of supersymmetry breaking mediation from brane intersections to the rest of the world-volume states, generating a mass. The result reproduces the expectations from the effective gauge theory, with the trivial inclusion of  Kaluza-Klein states. It exhibits at leading order in the expansion in powers  of $\frac{\bra D\ket}{M}$, where $M$ is the messenger mass scale, a tachyonic direction. This is expected \cite{Fayet:1978qc, Antoniadis:2006uj}, and is due to the form of couplings between the scalar adjoints and the messengers, as imposed by the original extended supersymmetry \cite{Benakli:2008pg}. 

Both parts are thus well under control and computable from the knowledge of the effective field theory. 
The issue that was  investigated in this work, and which needed an explicit check, is that there are no other contributions from the presence of the heavy string modes.

\section*{Acknowledgments}

This work was supported  in part by the European Commission under the ERC Advanced Grant 226371 and the contract PITN-GA-2009- 237920. I.A. was also supported in part by the CNRS grant GRC APIC PICS 3747. PA was supported by FWF P22000. PA would like to thank Marcus Berg and Robert Richter for discussions. MDG is supported by the German Science Foundation (DFG) under SFB 676.  MDG would like to thank the Centre de Physique Th\'eorique de l'Ecole Polytechnique, where part of this work was completed, for hospitality; and Emilian Dudas and Eran Palti for interesting discussions.

\appendix

\section{Field theory limit of  $N\approx 2$ amplitude}
\label{APPENDIX:FT}

Here we consider the field theory limit of the $N\approx2$ amplitude. This is instructive as it is useful for normalising the different parts of the amplitude, and also is interesting as it shall show how certain parts of the field theory diagram arise in the string amplitude.

\subsection{Field theory calculation}

The full two-point amplitude for the adjoints coupled to an $N=2$ hypermultiplet of mass $m$, with the scalar masses split by $\pm D$ is given by
\begin{align}
\C{A} ~=~& \C{A}_+ + \C{A}_- \nn\\
\C{A}_\pm ~=~&2g^2 \int \frac{d^4q}{(2\pi)^4} \frac{1}{q^2 - m^2 \mp D} + \frac{m}{q^2 - m^2 \mp D}\frac{m}{(q-p)^2 - m^2 \mp D} \nn\\
&~~\hskip 4cm - \frac{q^2 -  q \cdot p}{(q^2 - m^2)((q-p)^2 - m^2)}\nn\\
~=~& -i \frac{|\lambda|^2}{16\pi^2} \bigg[ - (m^2 \pm D) \log (m^2 \pm D) + m^2 \log m^2 \nn\\
&~~\hskip 1.5cm
- m^2 \log \frac{ - p^2 x(1-x) + m^2 \pm D}{ p^2 x(1-x) + m^2 } \nn\\
&~~\hskip 1.5cm + p^2 (1-x) \bigg\{  \frac{1}{\epsilon} + \log 4\pi + \gamma_E - \log [  m^2- p^2 x(1-x) ] \bigg\} \bigg].
\end{align}
The above can also be written in a friendlier format 
as
\begin{align}
\C{A}_\pm =& -i |\lambda|^2 \int \frac{d^4q}{(2\pi)^4} \int_0^1 dx -\frac{1}{q^2 + m^2 \pm D} +\frac{1}{q^2 + m^2 } +\frac{1}{2} \frac{ p^2 }{(q^2 - p^2 x(1-x) + m^2)^2}\nn\\
&\hskip 1.5cm - \frac{ m^2 }{(q^2 - p^2 x(1-x) + m^2 )^2} + \frac{ m^2 }{(q^2 - p^2 x(1-x) + m^2 \pm D)^2} .
\end{align}
This then yields
\begin{align}
\C{A} ~=~& -\frac{2 i g^2 p^2}{16\pi^2} \int \frac{dT}{T}  \int_0^1 dx e^{p^2 T x(1-x) -m^2 T} \nn\\
&~+ \frac{2i g^2}{16\pi^2} \int \frac{dT}{T^2} \exp [ -m^2 T] ( e^{DT/2} -  e^{-DT/2} )^2 \nn\\
&~- \frac{2 i g^2}{16\pi^2} \int \frac{dT}{T} m^2 ( e^{DT/2} -  e^{-DT/2} )^2 \int_0^1 dx e^{p^2 T x(1-x)}.
\end{align}
We shall recover this from a string computation.

\subsection{String Calculation}

Here we attempt to write the field theory limit of the string calculation, corresponding to $\A_{\Sigma^i\ov{\Sigma}^i}$ with $\theta_{ab}^i =0$; we can take the expressions from section \ref{SEC:STRINGY}. To compute the field theory limit, we consider only the $t\rightarrow \infty$ part of the amplitude, neglecting terms exponentially suppressed in $t$. In this limit, the function $Z_{N\approx 2}^{\theta_{ab}^i=0}$ becomes
\begin{equation}
Z_{N\approx 2}^{\theta_{ab}^i=0}~\rightarrow~ e^{- \pi \ap m^2 t} |I_{ab}| e^{3 \pi t/8} e^{-\pi |\theta_j| t/2}e^{-\pi |\theta_k| t/2} .
\end{equation}

\subsubsection{The contribution from worldsheet fermions}

Let us first deal with the worldsheet-fermionic contribution. Noting $\chi \rightarrow e^{-\pi \ap t k^2 (x - x^2)}$ we have
\begin{align}
\A_{\Sigma^i \ov{\Sigma}^i}^{\Psi_0} ~\rightarrow~& (-\mathrm{sign} (\theta_{ab}^j -\epsilon)\mathrm{sign} (\theta_{ab}^k -\epsilon)) i |I_{ab}| g^2 k^2\times\nn\\
&~~~~~~~~~~~~\int \frac{dt}{t} \frac{(2\pi)^2}{16\pi^4 } \int_0^{1} dx \exp [-\pi \ap k^2 (x - x^2)t - \pi \ap M_0^2 t -\pi \epsilon t]\nn\\
~&=~ i |I_{ab}| g^2 k^2\int \frac{dt}{t} \frac{1}{4\pi^2 } \int_0^{1} dx \exp [-\pi \ap k^2 (x - x^2)t - \pi \ap M_0^2 t].
\end{align}
Also 
\begin{align}
\A_{\Sigma^i \ov{\Sigma}^i}^{\Psi_1} =& - \frac{i}{2}|I_{ab}| g^2 k^2\int \frac{dt}{t} \frac{1}{4\pi^2 }  \int_0^{1} dx \exp [-\pi \ap k^2 (x - x^2)t - \pi \ap M_0^2 t].
\end{align}

Hence in total the fermionic contribution gives
\begin{align}
\A_{\Sigma^i \ov{\Sigma}^i}^{\Psi_0} + \A_{\Sigma^i \ov{\Sigma}^i}^{\Psi_1}~\rightarrow~ &\frac{i}{2}|I_{ab}| g^2 k^2\int \frac{dt}{t} \frac{1}{4\pi^2 }  \int_0^{1} dx \exp [-\pi \ap k^2 (x - x^2)t - \pi \ap M_0^2 t] \nn\\
~=~& \frac{i}{2}|I_{ab}| g^2 k^2\int \frac{dT}{T} \frac{1}{4\pi^2 }  \int_0^{1} dx \exp [- k^2 (x - x^2)T - M_0^2 T],
\end{align}
where $T \equiv \pi \ap t$.

\subsubsection{The bosonic contribution}

Now consider the bosonic piece. Here we find
\begin{align}
\A_{\Sigma^i \ov{\Sigma}^i}^X =& -\frac{g^2}{2} \int \frac{dt}{t^2} \frac{1}{16\pi^4 (\ap)^2} Z_{N\approx2}^{\theta_{ab}^i = 0}(t)  \frac{\vt (\epsilon it/2)}{\eta^3 (it/2)}  \vt (- \epsilon it/2) \\
&~~~~~~~~~~~~~~~~~~~~~~~~\times \prod_{\kappa=j,k} \vt ((\theta_{ab}^\kappa - \epsilon)it/2)\int_0^{it/2} dz \chi(z) 4 G (z) \nn\\
&\rightarrow~  -|I_{ab}| 2g^2 \int \frac{dt}{t^2} \frac{1}{16\pi^4 (\ap)^2} e^{- \pi \ap M_0^2 t} (e^{\pi \epsilon t/2}-e^{-\pi \epsilon t/2})^2   \int_0^{it/2} dz \chi(z)  G (z).\nn
\end{align}

Now recall
\begin{align}
\bra \partial X^j \partial \ov{X}^j\ket ~=~& \bra \partial X^j \partial \ov{X}^j\ket_{qu} + \bra \partial X^j \partial \ov{X}^j \ket_{cl} \nn\\
\langle \partial X (z) \partial \ov{X}(w) \rangle_{qu} ~=~& -\frac{\alpha'}{2} \partial_z\partial_w \log \theta_1 (z-w) .
\end{align}
The quantum part of the amplitude gives us
\begin{align}
\int_0^{it/2} dz \chi(z) &\langle \partial X (z) \partial \ov{X}(w) \rangle_{qu} \nn\\
~=~&  -\frac{\alpha'}{2} Z_{cl} \int_0^{it/2} dz \chi(z) \partial_z\partial_w \log \theta_1 (z-w) \nn\\
~=~& -\ap \pi i Z_{cl} + \frac{\alpha'}{2} \int_0^{it/2} dz \partial_z \chi \partial_w \log \theta_1 (z-w) \nn\\
~=~& -\ap \pi i Z_{cl} + i \pi^2 \alpha' (\ap k^2)  \int_0^{1} dx (1-2x) \exp [ \pi \ap t k^2 (x - x^2)] \nn\\
~&\rightarrow ~-\ap \pi i Z_{cl} , 
\end{align}
with no further contribution. It is interesting how the Feynman-parameter independent contribution arises here. The classical part gives
\begin{align}
\int_0^{it/2} dz \chi(z)& \bra \partial X^j \partial \ov{X}^j\ket_{cl} 
\nn\\
=&\sum_{n_j,m_j}  \int_0^1 dx \exp [ \pi \ap k^2 t(x - x^2)] \bigg[ \pi^2 it \bigg(  n_j^2 \frac{(\ap)^2}{\V_j^2} + \bigg( m_j \frac{T_2^j}{\V_j} + \frac{y}{2\pi}\bigg)^2 \bigg) \bigg] \nn\\
&\times \exp \bigg[ - \frac{\pi \ap t}{\V_j^2} \bigg| n_j  + i \bigg( m_j \frac{T_2^j}{\ap} + \frac{y \V_j}{2\pi \ap}\bigg) \bigg|^2 \bigg] \nn\\
=&\sum_{n_j,m_j} \int_0^1 dx \pi^2 it \ap m^2\bigg] \exp \bigg[\pi \ap k^2 t(x - x^2) - \pi \ap m^2 t\bigg].
\end{align}
Thus the total bosonic contribution is
\begin{align}
\A_{\Sigma^i \ov{\Sigma}^i}^X =& -|I_{ab}| 2g^2 \int dt \frac{1}{16\pi^4 (\ap)^2} e^{- \pi \ap m^2 t} (e^{\pi \epsilon t/2}-e^{-\pi \epsilon t/2})^2   \bigg[ \frac{-i\pi \ap}{t^2} + \frac{\pi^2 i \ap m^2 }{t}\bigg] \nn\\
=&~ i|I_{ab}| \frac{2g^2}{16\pi^2} \int dT e^{- m^2 T} (e^{DT/2}-e^{-DT/2})^2   \bigg[ \frac{1}{T^2} - \frac{m^2 }{T} e^{- k^2 T x(1-x)}\bigg].
\end{align}

\subsubsection{Total}

Then putting the whole amplitude together we have
\begin{align}
\C{A} ~=~& i|I_{ab}| \frac{2g^2}{16\pi^2} \int dT e^{- m^2 T} (e^{DT/2}-e^{-DT/2})^2   \bigg[ \frac{1}{T^2} - \frac{m^2 }{T} e^{- k^2 T x(1-x)}\bigg] \nn\\
&~~ + i|I_{ab}| \frac{2g^2}{16\pi^2} k^2\int \frac{dT}{T}  \int_0^{1} dx\;  e^{- k^2 (x - x^2)T -m^2 T}.
\end{align}
This exactly matches the field theory amplitude when we put $k^2 =-p^2$.

\end{document}